\documentclass[a4paper,12pt]{article}
\usepackage{amssymb}
\usepackage{amsmath}
\usepackage{ascmac}
\usepackage{comment}
\usepackage{wrapfig}
\usepackage{ulem}%波線
\usepackage{graphicx,color}
\newtheorem{thm}{Theorem}[subsection]
\newtheorem{prop}[thm]{Proposition}
\newtheorem{lem}[thm]{Lemma}

\newtheorem{rem}[thm]{Remark}

\def\C{{\mathbb C}}
\def\P{{\mathbb P}}

\def\dfrac#1#2{{\displaystyle\frac{#1}{#2}}}
\def\sfrac#1#2{{#1}/{#2}}
\def\o#1{{\overline{#1}}}
\def\u#1{{\underline{#1}}}
\def\ds{\displaystyle}
\def\prf{\noindent{\bf Proof.} }

\def\sq{\hfill$\square$}

\makeatletter
\@addtoreset{equation}{section}

\makeatother
\begin{document}

\begin{center}
{\Large {\bf 
%Three representations of affine Weyl  group symmetry of type $E_7^{(1)}$
Three $q$-Painlev\'e equations with affine Weyl group symmetry of type $E_7^{(1)}$
%assocciated with three realizations of the surface type $A_1^{(1)}$
}}

\vspace{10mm}
{\large By}

\vspace{5mm}
{\large Hidehito Nagao$^1$ and Yasuhiko Yamada$^2$}

(Akashi College$^1$ and Kobe University$^2$, Japan)
\end{center}

\noindent
{\bf Abstract.}
%It is well known that the $q$-Painlev\'e equations of affine Weyl  group symmetry type $E_7^{(1)}$  (i.e. $q$-$E_7^{(1)}$ equations) is derived by using degenerations of the $q$-$E_8^{(1)}$ 
%equation. 
Three $q$-Painlev\'e type equations are derived through degenerations of the $q$-Painlev\'e equation with affine Weyl group symmetry of type $q$-$E_8^{(1)}$.
%Three $q$-Painlev\'e equations of affine Weyl  group symmetry type $E_7^{(1)}$ (i.e. $q$-$E_7^{(1)}$ equations) have been derived as simple forms by using degenerations of the $q$-$E_8^{(1)}$ 
%$q$-$A_0^{(1)}$ 
%equation. 
%We study their realizations of surface type $A_1^{(1)}$ (i.e. singular point configuration) and their evolution directions and prove a relation among them in terms of affine Weyl group.
The three $q$-Painlev\'e type equations are associated with different realizations of 
%affine Weyl  group symmetry type $E_7^{(1)}$. 
the same symmetry/surface type $q$-$E_7^{(1)}$/$q$-$A_1^{(1)}$. 
We give three representations of affine Weyl  group actions of type $q$-$E_7^{(1)}$, and relations among them. The three $q$-$E_7^{(1)}$ equations are also derived from the three representations respectively.

\vspace{5mm}


%14H70: Relationships with integrable systems
%33D15: Basic hypergeometric functions in one variable, ${]_r\phi_s$
%33D70: Other basic hypergeometric functions and integrals in several variables
%34M55: Painlev\UTF{00E9} and other special equations; classification, hierarchies; isomonodromic deformations
%37K20: Relations with algebraic geometry, complex analysis, special functions
%39A13: Difference equations, scaling ($q$-differences)
%41A21: Pad\UTF{00E9} approximation

%\maketitle
%\tableofcontents 目次
\renewcommand\baselinestretch{1.2}
%%%%%%%%%%%%%%%%%%%
\section{Introduction}\label{sect:intro}

 In view of the Weyl group symmetry, the discrete Painlev\'e equations with higher symmetries such as $E_8^{(1)}$, $E_7^{(1)}$, etc. are very complicated in general.  Therefore we need some ideas to obtain their explicit forms (see for examples \cite{GR99, ORG01,Sakai01}). Namely, those equations 
admit
 simple expression only when we choose both the representations and the evolution directions properly.

Conversely,  for a given explicit equation, one needs some efforts to understand it from geometric or symmetric points of view.  
%E8 の簡単な形の方程式から退化させることによって３つの簡単な形のE７方程式が得られる
%これらの方程式のワイル群的な由来（解釈）や相互の関係を明らかにすることをこの論文の目的とする
In this paper, we study such problems in case of three $q$-$E_7^{(1)}$ equations arising from certain degenerations of the simple form of the $q$-$E_8^{(1)}$ equation \cite{KNY17, Yamada14}. 

The purpose of this paper is to work out the following studies.

\begin{itemize}
\item To derive the three $q$-$E_7^{(1)}$ equations as degenerations of the $q$-$E_8^{(1)}$ equation. 

\item To understand the obtained three 
$q$-$E_7^{(1)}$ equations in view of the Weyl group symmetry. 
%They have different realizations of the singular points configuration in $\P^1 \times \P^1$ and the corresponding different evolution directions respectively.
\item To clarify the relations among the three $q$-$E_7^{(1)}$ equations using the Weyl group symmetry.
%We give three representations of the affine Weyl group actions with symmetry type $q$-$E_7^{(1)}$ and relations among the three representations. The three $q$-$E_7^{(1)}$ equations are also derived from the three representations and the corresponding different evolution directions respectively.
\end{itemize}

This paper is organized as follows. In Section \ref{sec:ev}, we derive the three $q$-$E_7^{(1)}$ equations by using the degenerations of the $q$-$E_8^{(1)}$ equation. In Section \ref{sec:aW}, we give three representations of the affine Weyl group actions for the three $q$-$E_7^{(1)}$ equations without using the affine Weyl group symmetry, 
%with symmetry type $q$-$E_7^{(1)}$ 
and relations among the three representations. In Section \ref{sec:eq}, we derive the three equations from the three representations. In Appendix \ref{sec:ORG}, we show a relation between the $q$-$E_8^{(1)}$ equation given in \cite{KNY17, Yamada14} and ORG form \cite{ORG01}.  In Appendix \ref{sec:ev_rel}, we investigate a correspondence between two of the three $q$-$E_7^{(1)}$ equations. 
%\red{cases (i) and (iii)}. 
In Appendix \ref{sec:trans}, we give three translations on the root variables corresponding to the three representations. 

\section{Three $q$-$E_7^{(1)}$ equations with different realizations of surface type $q$-$A_1^{(1)}$}\label{sec:ev}

In this section we recall the $q$-$E_8^{(1)}$ equation 
%given in \cite{KNY17, Yamada14}, 
and will derive three $q$-$E_7^{(1)}$ equations by taking suitable degeneration limits.

Throughout this section, we use the following notations. Parameters $q$, $\kappa_1, \kappa_2, v_1, \ldots, v_8 \in\C^{\times}$ are complex parameters with a constraint 
\begin{equation}\label{eq:E8_Le_cons}
\kappa_1^2\kappa_2^2=q\prod_{i=1}^{8}v_i,
\end{equation} 
and $f, g$ be dependent variables.
Then we consider a time evolution $T$ given as
\begin{equation}\label{eq:ev_E8_te}
T : (v_1, \ldots, v_8, \kappa_1, \kappa_2, f, g) \mapsto (v_1, \ldots, v_8,  \frac{\kappa_1}{q}, q\kappa_2, \o{f}, \o{g}),
\end{equation} 
where 
%for any object $X$ the time evolution is denoted as
 $\o{X}=T(X)$ and $\u{X}=T^{-1}(X)$. 
 
%\subsection{Degeneration from $q$-$E_8^{(1)}$ equation }\label{subsec:ev_deg}
\subsection{$q$-$E_8^{(1)}$ equation}
The $q$-$E_8^{(1)}$ equation\footnote{The $q$-$E_8^{(1)}$ was firstly presented 
in \cite{ORG01} and the birational form is more complicated than one given in \cite{Yamada14}. See Appendix \ref{sec:ORG} for more details.}
\footnote{See \cite{KNY17, Yamada11,Yamada14} for the Lax equation, and \cite{KNY17, Yamada14} for hypergeometric special solutions of  (\ref{eq:ev_E8_ev}).} is given by (\ref{eq:ev_E8_te}) as the following simple form \cite{KNY17,Yamada14}:
\begin{equation}\label{eq:ev_E8_ev}
\begin{array}l
\ds\frac{\{g-G(\frac{\kappa_1}{z})\}\{\u{g}-\u{G}(\frac{\kappa_1}{z})\}}{\{g-G(z)\}\{\u{g}-\u{G}(z)\}}=\frac{U(\frac{\kappa_1}{z})}{U(z)},\quad {\rm for} \ f=F(z),
\\[5mm]
\ds\frac{\{f-F(\frac{\kappa_2}{z})\}\{\o{f}-\o{F}(\frac{\kappa_2}{z})\}}{\{f-F(z)\}\{\o{f}-\o{F}(z)\}}=\frac{U(\frac{\kappa_2}{z})}{U(z)},\quad {\rm for} \ g=G(z),
\end{array}
\end{equation}
 where
\begin{equation}\label{eq:ev_E8_FGU}
F(z)=z+\frac{\kappa_1}{z}, \quad G(z)=z+\frac{\kappa_2}{z}, \quad U(z)=z^{-4}\prod_{i=1}^{8}(z-v_i).
\end{equation}
%where
%\begin{equation}\label{eq:E8_Le_w}
%\begin{array}l
%\ds\frac{w\{f-F(z)\}\{\o{f}-\o{F}(z)\}}{U(z)}=\frac{\{g-G(\frac{\kappa_1}{z})\}\{g-G(\frac{\kappa_1}{qz})\}}{(z-\frac{\kappa_1}{z})(z-\frac{\kappa_1}{qz})}
%\\[5mm]
%\phantom{\ds\frac{w\{f-F(z)\}\{\o{f}-\o{F}(z)\}}{U(z)}}
%\ds =\frac{(\kappa_1-\kappa_2)(\kappa_1-q\kappa_2)}{\kappa_1^2} \quad \mbox{for} \quad g=G(z).
%\end{array}
%\end{equation}

\begin{figure}[htbp]
 \begin{minipage}{0.6\hsize}
 % \begin{center}
The eight singular points are  
\begin{equation}\label{eq:ev_E8_8p}
\ds(f,g)=\big(F(v_i), G(v_i)\big)_{i=1,\ldots,8}
\end{equation}
on a curve of bidegree $(2,2)$ with a node. This configuration is the surface type $q$-$A_0^{(1)}$ in Figure \ref{fig:E8}. %and unique.
%\end{center}
%where $F(z)=z+\dfrac{\kappa_1}{z}$ and $G(z)=z+\dfrac{\kappa_2}{z}$.
   \end{minipage}
 \begin{minipage}{0.4\hsize}
 \begin{center}
 \includegraphics[bb=100 520 300 750, scale=0.35]{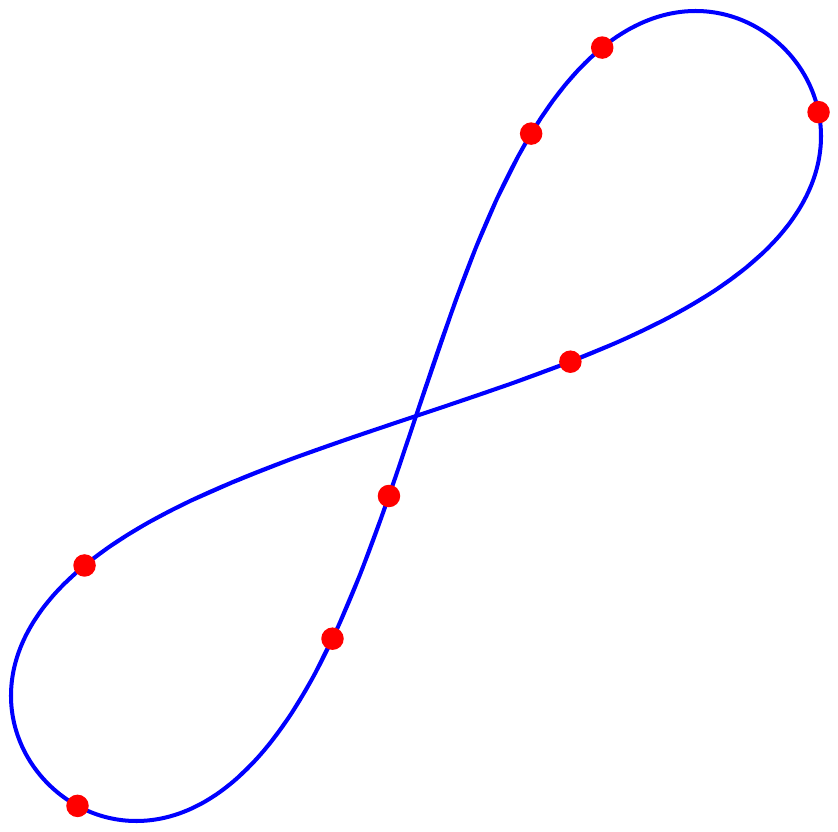}
  \caption{Surface type $q$-$A_0^{(1)}$}\label{fig:E8}
  \end{center}
\end{minipage}
\end{figure}

%\newpage
In next subsections, we will consider three degenerations of the $q$-$E_8^{(1)}$ equation (\ref{eq:ev_E8_ev}). 

\subsection{Degeneration 1}
Replacing parameters $\kappa_i$, $v_j$, $g$ by
\begin{equation}\label{eq:ev_E7_T1_deg}
\begin{array}l
\ds \kappa_i \to \varepsilon \kappa_i \ (i=1,2), \quad v_j \to \varepsilon v_j \ (i=1,\ldots,4), \quad g \to \frac{1}{g}, 
%\quad w \to \frac{w}{g}.
\end{array}
\end{equation}
and taking a degeneration limit $\varepsilon \to 0$, we obtain the well-known $q$-$E_7^{(1)}$ equation\footnote{See \cite{KNY17, Nagao15} for the Lax equation and hypergeometric special solutions of  (\ref{eq:ev_E7_T1_ev}).} \cite{GR99}:
\begin{equation}\label{eq:ev_E7_T1_ev}
\begin{array}l
\ds\frac{(fg-\frac{\kappa_1}{\kappa_2})(f\u{g}-\frac{q\kappa_1}{\kappa_2})}{(fg-1)(f\u{g}-1)}
=\frac{\prod_{i=1}^4(f-\frac{\kappa_1}{v_i})}
{\prod_{i=5}^8 (f-v_i)}, \\[5mm]
\ds\frac{(fg-\frac{\kappa_1}{\kappa_2})(\o{f}g-\frac{\kappa_1}{q\kappa_2})}{(fg-1)(\o{f}g-1)}
=\frac{\prod_{i=1}^4 (g-\frac{v_i}{\kappa_2})}{\prod_{i=5}^8 (g-\frac{1}{v_i})}.
\end{array}
\end{equation}
\begin{figure}[htbp]
\begin{minipage}{0.6\hsize}
%\begin{center}
The eight singular points are
\begin{equation}\label{eq:ev_E7_T1_8p}
(f,g)=\Big(\frac{\kappa_1}{v_i},\frac{v_i}{\kappa_2}\Big)_{i=1}^4, \Big(v_i,\frac{1}{v_i}\Big)_{i=5}^8, 
\end{equation}
on the product of two curves of bidegree $(1,1)$ and $(1,1)$. This configuration for surface type $q$-$A_1^{(1)}$ %which we call Type 1 
is given in Figure \ref{fig:E7-1}. We call it configuration type 1.
%well-known for the realization of the surface type $q$-$A_1^{(1)}$  in Figure \ref{fig:E7-1}.
%\end{center}
 % \caption{一つめの図}
  %\label{fig:one}
 \end{minipage}
 \begin{minipage}{0.4\hsize}
 \begin{center}
 \includegraphics[bb=100 700 300 650, scale=0.4]{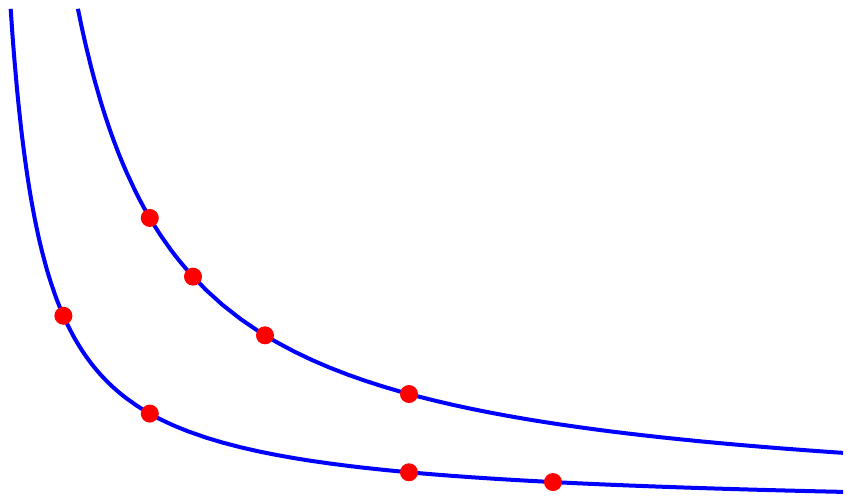}
   \end{center}
   \vspace{5mm}
   \caption{Configuration type 1}\label{fig:E7-1}
\end{minipage}
\end{figure}

\subsection{Degeneration 2}
Similarly, doing a replacement of parameters
\begin{equation}\label{eq:ev_E7_T2_deg}
\begin{array}l
\ds \kappa_1 \to \varepsilon \kappa_1, \quad v_i \to \varepsilon v_i  \ (i=1,2), 
%\quad w \to \frac{q \kappa_2^2 w}{\kappa_1^2 \varepsilon^2}, 
%\\[5mm]
%\ds y(z) \to H(z)y(z), \quad \o{y}(z) \to \o{H}(z)\o{y}(z), \quad \frac{H(qz)}{H(z)}=\frac{\kappa_1\varepsilon}{\kappa_2}, \quad \frac{\o{H}(z)}{H(z)}=1.
\end{array}
\end{equation}
and taking a degeneration limit $\varepsilon \to 0$, one gets the $q$-$E_7^{(1)}$ equation\footnote{In \cite{Nagao17-2} the Lax equation and hypergeometric special solutions of (\ref{eq:ev_E7_T2_ev}) are given, and a relation between the Lax equations for (\ref{eq:ev_E7_T1_ev}) and (\ref{eq:ev_E7_T2_ev}) is also shown.} :
\begin{equation}\label{eq:ev_E7_T2_ev}
\begin{array}l
\ds\{g-G(f)\}\{\u{g}-\u{G}(f)\}=\frac{\prod_{i=3}^8 (f-v_i)}{f^2\prod_{i=1}^2(f-\frac{\kappa_1}{v_i})},
\\[5mm]
\ds\frac{(f-\frac{\kappa_2}{z_s})(\o{f}-\frac{\kappa_2}{z_s})}{(f-z_s)(\o{f}-z_s)}=\frac{z_s^2}{(\frac{\kappa_2}{z_s})^2}\frac{\prod_{i=3}^8 (\frac{\kappa_2}{z_s}-v_i)}{\prod_{i=3}^8 (z_s-v_i)}, \quad {\rm for} \ g=G(z_s).
\end{array}
\end{equation}
\begin{figure}[htbp]
 \begin{minipage}{0.6\hsize}
%  \begin{center}
This has been regarded in \cite{Nagao17-2, NY18-1, NY18-2} as a variation of the $q$-$E_7^{(1)}$ equation (\ref{eq:ev_E7_T1_ev}). The eight singular points are 
\begin{equation}\label{eq:ev_E7_T2_8p}
(f,g)=\big({v_i}, G(v_i)\big)_{i=3}^8,\left(\frac{\kappa_1}{v_i},\infty\right)_{i=1,2}, 
\end{equation}
on the product of a curve and line of bidegree $(2,1)$ and $(0,1)$. This configuration for the surface type $q$-$A_1^{(1)}$ is given  in Figure \ref{fig:E7-2}. 
We call it configuration type 2.
% and given in \cite{Nagao17-2, NY18-1, NY18-2}.
%\end{center}
%where $G(z)=v_i+\dfrac{\kappa_2}{v_i}$.
 % \caption{一つめの図}
  %\label{fig:one}
 \end{minipage}
 \begin{minipage}{0.4\hsize}
 \begin{center}
 \includegraphics[bb=100 450 300 700, scale=0.3]{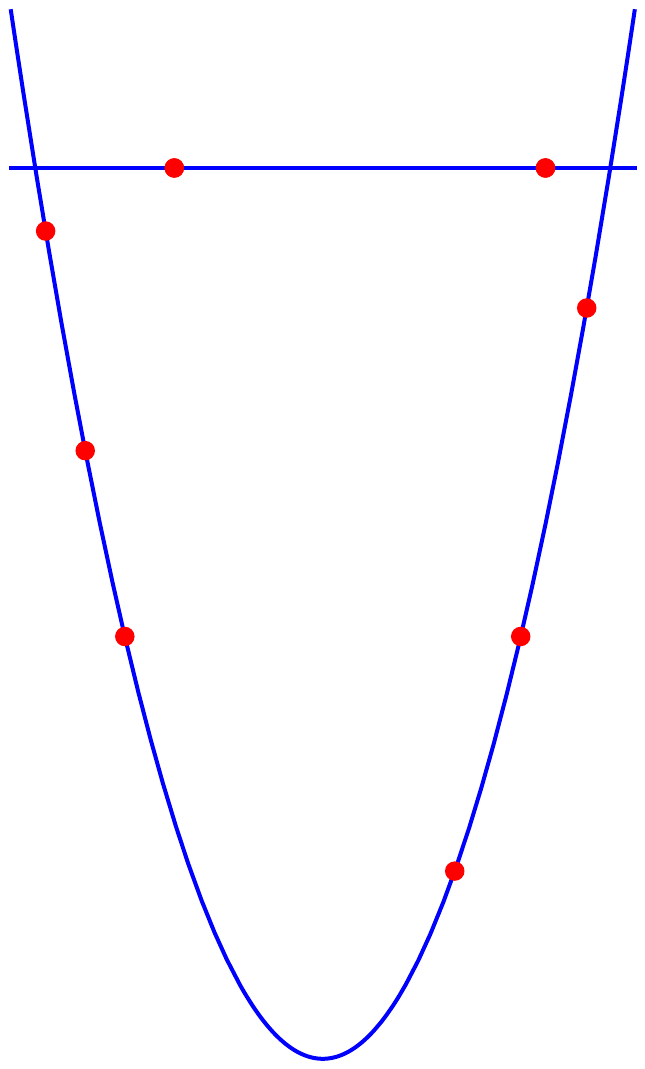}
   \caption{Configuration type 2}\label{fig:E7-2}
  \end{center}
\end{minipage}
\end{figure}

\vspace{5mm}
%\red{
%\begin{rem}\label{rem:ev_rel_Lax}
%It was in \cite[Remark 4.1]{NY18-2} shown that the Lax equations for each evolution equations (\ref{eq:ev_E7_T1_ev}) and (\ref{eq:ev_E7_T2_ev}) are equivalent under a relation between independent variables. 
%\end{rem}
%}
\subsection{Degeneration 3}
Doing a replacement of parameters
\begin{equation}\label{eq:ev_E7_T3_deg}
v_1\to \varepsilon v_1,\quad v_2 \to \varepsilon^{-1} v_2, 
%\quad  -w \to \frac{v_2 w}{\varepsilon}.
\end{equation}
and applying a degeneration limit $\varepsilon \to 0$, 
we have the following novel evolution equation:
\begin{equation}\label{eq:ev_E7_T3_ev}
\begin{array}l
\ds\frac{\{g-G(\frac{\kappa_1}{z})\}\{\u{g}-\u{G}(\frac{\kappa_1}{z})\}}{\{g-G(z)\}\{\u{g}-\u{G}(z)\}}=\frac{U(\frac{\kappa_1}{z})}{U(z)},\quad {\rm for} \quad f=F(z),
\\[5mm]
\ds\frac{\{f-F(\frac{\kappa_2}{z})\}\{\o{f}-\o{F}(\frac{\kappa_2}{z})\}}{\{f-F(z)\}\{\o{f}-\o{F}(z)\}}=\frac{U(\frac{\kappa_2}{z})}{U(z)},\quad {\rm for} \quad g=G(z),
\end{array}
\end{equation}
\begin{figure}[htbp]
 \begin{minipage}{0.61\hsize}
%  \begin{center}
where $F(z)$ and $G(z)$ are as in (\ref{eq:ev_E8_FGU}) but only $U(z)$ (\ref{eq:ev_E8_FGU}) is changed to $U(z)=z^{-3}\prod_{i=3}^8(z-v_i)$. This equation (\ref{eq:ev_E7_T3_ev}) is a kind of the $q$-$E_7^{(1)}$ equation. 
%since it has the surface type $q$-$A_1^{(1)}$ (see the eight singular points (\ref{eq:ev_E7_T3_8p})) and the point configuration in Figure \ref{fig:E7-3}).
The eight singular points are \begin{equation}\label{eq:ev_E7_T3_8p}
\begin{array}l
(f,g)=\big(F(v_i), G(v_i)\big)_{i=3}^8, \Big(\frac{1}{\varepsilon}, 
\frac{1-v_1v_2/\kappa_1}{(1-v_1v_2/\kappa_2)\varepsilon}\Big)_2, 
\end{array}
\end{equation} 
%See two methods \cite[Section 4.3]{KNY17}) and \cite[Section 7.1]{KNY17} for finding the blow-up (i.e. the gradient) on the point $(\infty, \infty)$
%\end{center}
%where $F(z)=z+\dfrac{\kappa_1}{z}$ and $G(z)=z+\dfrac{\kappa_2}{z}$.
 % \caption{一つめの図}
  %\label{fig:one}
 \end{minipage}
 \begin{minipage}{0.4\hsize}
 \begin{center}
 \includegraphics[bb=100 520 300 700, scale=0.35]{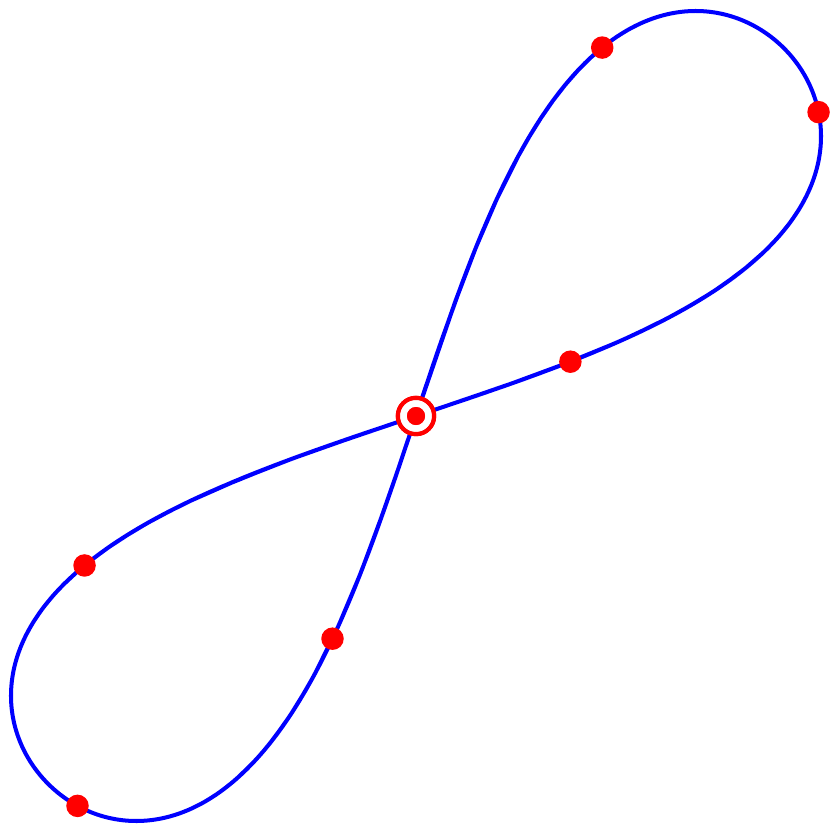}
  \caption{Configuration type 3}\label{fig:E7-3}
  \end{center}
\end{minipage}
\end{figure}  

\noindent
on a curve of bidegree $(2,2)$ with a node. Here, the last point is a double point at $(\infty, \infty)$ with the limit  
%\footnote{See 
%%Appendix \ref{sec:method} 
%\cite[Section 4.3]{KNY17} 
%%(also \cite[Section 7.1]{KNY17}) 
%for a convenient method 
%to find the gradient on $(\infty, \infty)$ instead of blow-ups.} 
$\lim_{\varepsilon\to 0} \frac{g}{f}=\frac{1-v_1v_2/\kappa_1}{1-v_1v_2/\kappa_2}$. This configuration for the surface type $q$-$A_1^{(1)}$ is given in Figure \ref{fig:E7-3}, and it is not considered so far. We call it configuration type 3. 

\begin{rem}
The equations (\ref{eq:ev_E7_T1_ev}) and (\ref{eq:ev_E7_T3_ev}) are equivalent with each other (see Appendix \ref{sec:ev_rel}). 
\end{rem}

%\begin{rem}
%Instead of the blow-up on the point $(\infty, \infty)$, there exist some practical methods for finding the point of indeterminacy of the bi-rational mapping. One is that sufficient number of iterations give all the point of indeterminacy (see \cite[Section 4.3]{KNY17}). Another is that the Lax equation can be regarded as the curve of bi-degree $(3,2)$ in $(f,g)$ and the curve is uniquely determined  up to a scalar multiple by the vanishing property at the passing twelve points including the 8 singular points (see \cite[Section 4.3]{KNY17}).) 
%\end{rem}
 %%%%%%%%%%%%%%%%%%%%%%%%
\section{Three representations of affine Weyl group actions 
% and the corresponding $q$-$E_7^{(1)}$ equations
}
\label{sec:aW}
Let us consider the affine Weyl group of type $E_7^{(1)}$ 
%(simple reflections) 
$\langle s_0, \ldots s_7 \rangle$, and 
%(lattice isomorphisms) 
the Dynkin diagram automorphism $\pi$ satisfying the fundamental relations
\begin{equation}\label{eq:aW_rel}
\begin{array}l
s_i^2=1 \quad \mbox{for \ all} \ i, \\[2mm]
s_is_j=s_js_i \quad \mbox{when} \ (a_{ij}, a_{ji})=(0,0), \\[2mm]
s_is_js_i=s_js_is_j \quad \mbox{when} \ (a_{ij}, a_{ji})=(-1,-1), \\[2mm]
\pi^2=1, \quad \pi s_i=s_i \pi \ (i=0,4), \quad \pi s_1s_6=s_2s_7\pi, \quad \pi s_2s_5=s_3s_6\pi, 
\end{array}
\end{equation}
for the generalized Cartan matrix of type $E_7^{(1)}$ 
\begin{equation}\label{eq:aW_E7_T1_cartan}
A=(a_{ij})=\begin{bmatrix}
 2 &    &    &    &-1 &    &    &     \\
    & 2 &-1 &    &    &    &    &      \\
    &-1 & 2 &-1 &    &    &    &     \\ 
    &    &-1 & 2 &-1 &    &    &      \\
-1 &    &    &-1 & 2 &-1 &    &      \\
    &    &    &    &-1 & 2 &-1 &       \\ 
    &    &    &    &    &-1 & 2 &-1     \\
    &    &    &    &    &    &-1 & 2 
 \end{bmatrix}.
\end{equation} 
In this section, we will give three representations of the affine Weyl group 
%actions 
$\langle s_0, s_1, \ldots s_7, \pi \rangle$, and relations among the three representations.
In this section, we use parameters $e_1, \ldots, e_8, h_1, h_2$ suitable for the description of the Weyl group symmetry instead of parameters $\kappa_1, \kappa_2, v_1, \ldots, v_8$. Their relations will be given case by case (see for examples (\ref{eq:aW_E7_T1_ve}), (\ref{eq:aW_E7_T2_ve}), (\ref{eq:aW_E7_T3_ve})).
In order to distinguish three representations of the affine Weyl group actions $s_i$ and $\pi$, 
% the simple reflections $s_i$ and the Dynkin diagram automorphisms $\pi$ for the roots %lattices 
%$\alpha_i$ in (\ref{eq:aW_E7_T1_root}) , (\ref{eq:aW_E7_T2_root})  and (\ref{eq:aW_E7_T3_root}) by
we make use of the notations $s_i^{[m]}, \pi^{[m]} \ (m=1,2,3)$. The actions $s_i^{[m]}, \pi^{[m]}$ are expressed by the reflections 
%\red{on parameters $h_1$, $h_2$, $e_1$, \ldots, $e_8$}.  
\begin{equation}\label{eq:aW_ref}
\begin{array}{rl}
r_{ij}%=r_{E_i-E_j}
:&e_i \leftrightarrow e_j,\\[5mm]
c%=r_{H_1-H_2}
:&h_1 \leftrightarrow h_2,\\[5mm]
\mu_{ij}
%=r_{H_2-E_i-E_j}
:& e_i \rightarrow \dfrac{h_2}{e_j}, \quad e_j \rightarrow \dfrac{h_2}{e_i}, \quad h_1 \rightarrow \dfrac{h_1h_2}{e_ie_j},\\[5mm]
\nu_{ij}
%=r_{H_1-E_i-E_j}
:& e_i \rightarrow \dfrac{h_1}{e_j}, \quad e_j \rightarrow \dfrac{h_1}{e_i}, \quad h_2 \rightarrow \dfrac{h_1h_2}{e_ie_j}.
%%\\[5mm]
%%s_{H_1+H_2-E_i-E_j-E_k-E_l}:& e_i \rightarrow \dfrac{h_1h_2}{e_je_ke_l}, \quad e_j \rightarrow \dfrac{h_1h_2}{e_ie_ke_l}, \quad e_k \rightarrow \dfrac{h_1h_2}{e_ie_je_l}, \quad e_l \rightarrow \dfrac{h_1h_2}{e_ie_je_k}, \\[5mm]
%%& h_1 \rightarrow \dfrac{h_1^2h_2}{e_ie_je_ke_l},\quad h_2 \rightarrow \dfrac{h_1h_2^2}{e_ie_je_ke_l}.
\end{array}
\end{equation} 

%We use the composition of the affine Weyl group actions $s_i$ by the notation
%\begin{equation}\label{eq:aW_simple}
%\begin{array}l
%s_{i_0i_1\ldots i_n}:=s_{i_0}s_{i_1}\cdots s_{i_n}, %\\[3mm]
%%s_{i_0i_1\ldots i_n}^{{m}}:=s_{i_0}^{{m}}s_{i_1}^{{m}}\cdots s_{i_n}^{{m}}, \ (m=1,2,3).
%\end{array}
%\end{equation} 
%%%%%%%%%%%%%%%%%%%%%
\subsection{Representation $1$ of affine Weyl group actions
%corresponding to (\ref{eq:ev_E7_T1_ev})
}\label{subsec:aW_E7_T1} 

%{\color{red}
%\begin{screen}
%回転$\pi$は退化では得られないが，その構成法は？Root dataの$\delta$の退化は？
%\end{screen}
%} 

%Applying a degeneration limit
%\begin{equation}\label{eq:aW_E7_T1_deg}
%\begin{array}l
%\ds h_i \to \varepsilon h_i (i=1,2), \quad e_j \to \varepsilon e_j (i=1,\ldots,4), \quad y \to y^{-1}.
%\end{array}
%\end{equation}
%into datas given in \S\ref{subsec:aW_E8}, we obtain the following datas.

%The well-known $q$-Painlev\'e equation with the symmetry type $E_7^{(1)}$ 

%It has been shown in \cite{KNY17} that 
%The $q$-$E_7^{(1)}$  equation (\ref{eq:ev_E7_T1_ev}) is given as the following representation of the affine Weyl group.

We consider $s_i^{[1]}$ and $\pi^{[1]}$ as  representation $1$ of the affine Weyl group actions $s_i$ and $\pi$.
The actions on parameters $e_1, \ldots, e_8, h_1, h_2$ are
\begin{equation}\label{eq:aW_E7_T1_para}
\begin{array}l
s_0^{[1]}=c, \quad s_1^{[1]}=r_{12}, \quad s_2^{[1]}=r_{23}, \quad s_3^{[1]}=r_{34}, \\
s_4^{[1]}=\mu_{45}, \quad s_5^{[1]}=r_{56}, \quad s_6^{[1]}=r_{67}, \quad s_7^{[1]}=r_{78}, \\
\pi^{[1]} : h_i \leftrightarrow h_i, \quad e_i \leftrightarrow e_{\sigma(i)}, \quad 
\sigma=\left(\substack{12345678\\87654321}\right).
\end{array}
\end{equation}
The actions on dependent variables $x, y$ are 
\begin{equation}\label{eq:aW_E7_T1_vari}
\begin{array}l
s_0^{[1]}: x \rightarrow \dfrac{1}{y}, \ y \rightarrow \dfrac{1}{x}, \\[5mm]
s_4^{[1]}: x \rightarrow \dfrac{e_4(h_1-h_2)x+h_2(e_4e_5-h_1)xy-h_1(e_4e_5-h_2)}{e_4(e_4e_5-h_2)xy-e_4(e_4e_5-h_1)-e_4e_5(h_1-h_2)y}, \\[5mm]
\pi^{[1]}: x \rightarrow \dfrac{h_1}{x}, \quad y \rightarrow \dfrac{1}{h_2y}. 
\end{array}
\end{equation}

The actions 
%$s_i^{(1)}$, $\pi^{(1)}$ 
(\ref{eq:aW_E7_T1_para}) and (\ref{eq:aW_E7_T1_vari}) is equivalent to the actions given in \cite[Section 8.4.2]{KNY17} up to trivial changes of parameters and variables. 

\subsection{Representation $2$ of affine Weyl group actions}\label{subsec:aW_E7_T2} 
We consider $s_i^{[2]}$ and $\pi^{[2]}$ defined by 
%as representation $2$ of the affine Weyl group actions $s_i$ and $\pi$. 
%Supposed that the actions $s_i^{[1]}$, $\pi^{[1]}$ and $s_i^{[2]}$, $\pi^{[2]}$ are related by a transformation 
\begin{equation}\label{eq:trans12}
s_i^{[2]}=\varphi_{12}s_i^{[1]}\varphi_{12}^{-1}, \quad \pi^{[2]}=\varphi_{12}\pi^{[1]}\varphi_{12}^{-1},
\end{equation}
where the action of $\varphi_{12}$ 
%$: (e_i, h_j, x, y)$ in (\ref{eq:aW_E7_T1_8p}) and (\ref{eq:aW_E7_T1_root}) $\to$  $(e_i, h_j, x, y)$ in   (\ref{eq:aW_E7_T2_8p}) and (\ref{eq:aW_E7_T2_root}) 
is given by
\begin{equation}\label{eq:ch12}
\begin{array}l
\ds\varphi_{12}: x\to \frac{h_2
   \left\{(e_3+e_4)x-e_3
   e_4+h_2-x y\right\}}{(e_3+e_4)
   h_2+(e_3 e_4 -h_2 )x-e_3
   e_4 y},\quad y\to
   \frac{1}{x}, 
   \\[8mm]\phantom{\ds\varphi_{12}=}
 \ds  e_3\to
   \frac{h_1}{e_3},\quad e_4\to
   \frac{h_1}{e_4},\quad h_1\to
   \frac{h_1 h_2}{e_3
   e_4},\quad h_2\to h_1.
\end{array}
\end{equation}
Then one can obtain the actions
\begin{equation}\label{eq:aW_E7_T2_para}
\begin{array}l
s_0^{[2]}=\mu_{34}, \quad s_1^{[2]}=r_{12}, \quad s_2^{[2]}=\nu_{23}, \quad s_3^{[2]}=r_{34}, \\
s_4^{[2]}=r_{45}, \quad s_5^{[2]}=r_{56}, \quad s_6^{[2]}=r_{67}, \quad s_7^{[2]}=r_{78}, \\[3mm]
\pi^{[2]}: h_2 \leftrightarrow \dfrac{h_1^2h_2}{e_3e_4e_5e_6}, %\\[3mm]\phantom{s_8^{[2]}=\pi:} 
e_1 \leftrightarrow e_8, \quad e_2 \leftrightarrow e_7,\quad e_i \leftrightarrow \dfrac{e_{\sigma(i)}}{h_1}, \quad 
\sigma=\left(\substack{3456\\6543}\right),
\end{array}
\end{equation}
% The action on dependent variables $x$ and $y$ is constructed as:
%Actions on dependent variables $x$ and $y$:
and
\begin{equation}\label{eq:aW_E7_T2_vari}
\begin{array}l
s_0^{[2]}: x \rightarrow \dfrac{h_2(xy-a_1x+a_2-h_2)}{(h_2-a_2)x+a_2y-a_1h_2}, \\[5mm]
s_2^{[2]}: y \rightarrow \dfrac{e_2e_3xy-e_3(e_2e_3-h_1)x-e_3h_1y-h_2(e_2e_3-h_1)}{e_2e_3(x-e_3)}, \\[5mm]
\pi^{[2]}: x \rightarrow \dfrac{h_1}{x},
%\\[5mm]\phantom{s_8^{[2]}=\pi: }
\ y \rightarrow \dfrac{h_1\{x^2yh_2-h_2b_1x^2-
(b_4-h_2b_2+h_2^2)x+b_4y-h_2b_3\}}{b_4(xy-x^2-h_2)},
 \end{array}
\end{equation}
where $a_i$ (resp. $b_i$) is given by $\prod_{i=3}^4(1+te_i)=\sum_{i=0}^2 a_it^i$ (resp.  $\prod_{i=3}^6(1+te_i)=\sum_{i=0}^4 b_it^i$). 
%Hence, the actions $s_i^{(1)}$ and $\pi^{(1)}$ in (\ref{eq:aW_E7_T1_para}) and $s_i^{[2]}$ and $\pi^{[2]}$ in (\ref{eq:aW_E7_T2_para}) are equivalent with each other under the transformation (\ref{eq:trans12}). 

\subsection{Representation $3$ of affine Weyl group actions}\label{subsec:aW_E7_T3} 
We consider $s_i^{[3]}$ and $\pi^{[3]}$ defined by 
% as representation $3$ of the affine Weyl group actions $s_i$ and $\pi$. 
%Supposed that the actions $s_i^{[2]}$, $\pi^{[2]}$ and $s_i^{[3]}$, $\pi^{[3]}$ are related by a transformation 
\begin{equation}\label{eq:trans23}
s_i^{[3]}=\varphi_{23}s_i^{[2]}\varphi_{23}^{-1}, \quad \pi^{[3]}=\varphi_{23}\pi^{[2]}\varphi_{23}^{-1}, 
\end{equation}
where the action of $\varphi_{23}$ is given by
\begin{equation}\label{eq:ch23}
\begin{array}l
\ds\varphi_{23}: x\to \frac{e_3 (h_1-
   h_2)+h_2 x-h_1 y}{e_3
   x-e_3 y-h_1+h_2},\quad y\to
   x, 
   \\[5mm]\phantom{\ds\varphi_{23}: }
   \ds e_2\to
   \frac{h_1}{e_2},\quad e_3\to
   \frac{h_1}{e_3},\quad h_1\to
   \frac{h_1 h_2}{e_2
   e_3},\quad h_2\to h_1.
   \end{array}
\end{equation}
Then one can get the actions 
\begin{equation}\label{eq:aW_E7_T3_para}
\begin{array}l
s_0^{[3]}=r_{34}, \quad s_1^{[3]}=\nu_{12}, \quad s_2^{[3]}=c, \quad s_3^{[3]}=\nu_{34},\\
s_4^{[3]}=r_{45}, \quad s_5^{[3]}=r_{56}, \quad s_6^{[3]}=r_{67}, \quad s_7^{[3]}=r_{78}, \\[3mm]
\pi^{[3]}: h_1 \rightarrow \dfrac{e_2^2e_3e_4e_5e_6}{h_1^2h_2^2}, \quad h_2 \rightarrow \dfrac{e_2^2e_3e_4e_5e_7}{h_1^2h_2^2}, \quad e_1 \leftrightarrow e_8, \quad e_2 \rightarrow  \dfrac{h_1^2h_2^2}{e_2^2e_3e_4e_5e_6e_7}, 
\\[3mm]\phantom{\pi:} 
e_3 \rightarrow \dfrac{h_1h_2}{e_2e_4e_5}, \quad e_4 \rightarrow \dfrac{h_1h_2}{e_2e_3e_5}, \quad e_5 \rightarrow \dfrac{h_1h_2}{e_2e_3e_4}, \quad e_6 \rightarrow \dfrac{h_2}{e_2}, \quad e_7 \rightarrow \dfrac{h_1}{e_2},
\end{array}
\end{equation}
and
\begin{equation}\label{eq:aW_E7_T3_vari}
\begin{array}l
s_1^{[3]}: y \rightarrow \dfrac{-h_2(e_1e_2-h_1)x+h_1(e_1e_2-h_2)y}{e_1e_2 h_{-}}, \quad s_2^{[3]}: x \leftrightarrow y, \\
s_3^{[3]}: y \rightarrow \dfrac{a_2 h_{-} xy+a_1h_2(a_2-h_1)x-a_1h_1(a_2-h_2)y+(a_2-h_1)(a_2-h_2) h_{-}}{a_2\{(a_2-h_2)x-(a_2-h_1)y-a_1h_{-}\}}, \\[3mm]
\pi^{[3]}: x \rightarrow \Phi, \quad 
y \rightarrow \Phi |_{e_6 \to e_7}.
\end{array}
\end{equation}
Here
\begin{equation}
%\begin{array}l
\Phi=\dfrac{h_1h_2\left\{(b_4(x-y)-h_{-}b_3)(x-y)+(h_2x-h_1y+h_{-}b_1)(h_2x-h_1y)+b_2(h_{-})^2\right\}
}{e_2b_4\{h_2x^2+h_1y^2-h_{+} xy+(h_{-})^2\}},
%\dfrac{1}{h_1h_2^2x^2-h_1h_2(h_1+h_2)xy+h_1^2h_2y^2+h_1h_2(h_1-h_2)^2} \\[2mm]
%\phantom{\Phi}
%\times [e_2(b_0+h_2^2)x^2-2e_2(b_0+h_1h_2)xy
%-e_2(h_1-h_2)(b_1-h_2b_3)x \\[2mm]
%\phantom{\Phi=}
%+e_2(b_0+h_1^2)y^2+e_2(h_1-h_2)(b_1-h_1b_3)y+e_2b_2(h_1-h_2)^2],
%\end{array}
\end{equation}
and $h_{\pm}$ mean $h_1\pm h_2$ respectively, and $a_i$, $b_i$ are gievn below (\ref{eq:aW_E7_T2_vari}). 
%Therefore, the actions $s_i^{[2]}$ and $\pi^{[2]}$ in (\ref{eq:aW_E7_T2_para}) and $s_i^{[3]}$ and $\pi^{[3]}$ in (\ref{eq:aW_E7_T3_para}) are equivalent with each other under the transformation (\ref{eq:trans23}). 
 
\section{Three $q$-$E_7^{(1)}$ equations from affine Weyl group actions}\label{sec:eq}
In this section, the $q$-$E_7^{(1)}$ equations (\ref{eq:ev_E7_T1_ev}), (\ref{eq:ev_E7_T2_ev}), (\ref{eq:ev_E7_T3_ev}) will be derived from the three representations $1, 2, 3$ 
%(\ref{eq:aW_E7_T1_para}), (\ref{eq:aW_E7_T1_vari}), (\ref{eq:aW_E7_T2_para}), (\ref{eq:aW_E7_T2_vari}), (\ref{eq:aW_E7_T3_para}), (\ref{eq:aW_E7_T3_vari}) 
of affine Weyl  group actions in Section \ref{sec:aW}.
%$s_0, \ldots, s_7, \pi$ with symmetry type $q$-$E_7^{(1)}$ respectively.

\subsection{The $q$-$E_7^{(1)}$ equation from representation $1$}\label{sec:eq_E7_T1}
Let us define the following composition of the Weyl group elements $s_0, \ldots, s_7$ as
%Under the composed action 
\begin{equation}\label{eq:aW_E7_T1_translation} 
\begin{array}l
%T_1=(s_0^{(1)} s_{4567345623451234}^{(1)})^2,
T_1=(s_{04567345623451234})^2, 
\end{array}
\end{equation}
where $s_{i_0i_1\ldots i_n}:=s_{i_0}s_{i_1}\cdots s_{i_n}$.
Then the action $T_1^{[1]}$ of $T_1$ in representation $1$ is expressed as %the root variables
%$\alpha_i$ in (\ref{eq:aW_E7_T1_root}) is transformed by
\begin{equation}\label{eq:aW_E7_T1_translation2}
\begin{array}l
T_1^{[1]}=(s_{04567345623451234}^{[1]})^2, 
\end{array}
\end{equation}
and we have
\begin{prop}
The action $T_1^{[1]}$ (\ref{eq:aW_E7_T1_translation2}) 
%on the root 
gives the following 
%birational action of $x$, $y$:
equation
\begin{equation}\label{eq:aW_E7_T1_br}
\begin{array}l
\dfrac{{T_1^{(1)}}^{-1}(x)y-1}{{T_1^{(1)}}^{-1}(x)y-\frac{qh_2}{h_1}}\dfrac{xy-\frac{h_1}{h_2}}{xy-1}=\dfrac{\prod_{i=1}^4(y-\frac{e_i}{h_2})}{\prod_{i=5}^8(y-\frac{1}{e_i})},\\[5mm]
\dfrac{xT_1^{(1)}(y)-1}{\frac{qh_1}{h_2}xT_1^{(1)}(y)-1}\dfrac{\frac{h_2}{h_1}xy-1}{xy-1}=\dfrac{\prod_{i=1}^4(\frac{e_ix}{h_1}-1)}{\prod_{i=5}^8(\frac{x}{e_i}-1)}. 
\end{array}
\end{equation}
Here the equation (\ref{eq:aW_E7_T1_br}) is equivalent to the $q$-$E_7^{(1)}$ equation (\ref{eq:ev_E7_T1_ev}), and it has the singular point configuration (i.e. configuration type $1$ for surface type $q$-$A_1^{(1)}$ in Figure \ref{fig:E7-1})
\begin{equation}\label{eq:aW_E7_T1_8p}
(x,y)=\Big(\frac{h_1}{e_i},\frac{e_i}{h_2}\Big)_{i=1,\ldots,4}, \Big(e_i,\frac{1}{e_i}\Big)_{i=5,\ldots, 8}.
\end{equation}
\end{prop}

\prf
Firstly, from the actions $s_i^{[1]}$ (\ref{eq:aW_E7_T1_para}) and (\ref{eq:aW_E7_T1_vari}), the action $T_1^{[1]}$ (\ref{eq:aW_E7_T1_translation2}) can be rewritten into the equation (\ref{eq:aW_E7_T1_br}). 

Next, under the transformation
%\footnote{see \cite{KNY17} for elliptic, multiplicative and additive cases of the symmetry type $E_8^{[1]}$ (i.e $e$-$E_8^{[1]}$, $q$-$E_8^{[1]}$, $d$-$E_8^{[1]}$).}
\begin{equation}\label{eq:aW_E7_T1_ve}
\begin{array}l
\ds v_i=\frac{e_i}{\lambda}, \quad \kappa_1=\frac{h_2}{\lambda^2}, \quad \kappa_2=\frac{h_1}{\lambda^2}, \quad f=\frac{1}{\lambda y}, \quad g=\frac{\lambda}{x}, 
\end{array}
\end{equation}
where ${T_1^{[1]}}^{-1}(h_1)=\frac{qh_2^2}{h_1}$, $T_1^{[1]}(h_2)=\frac{qh_1^2}{h_2}$, $T_1^{[1]}(\lambda)=\frac{\lambda T_1^{[1]}(h_2)}{h_1}$, 
the equation (\ref{eq:aW_E7_T1_br}) is equivalent to the $q$-$E_7^{[1]}$ equation (\ref{eq:ev_E7_T1_ev}), and it has the configuration (\ref{eq:aW_E7_T1_8p}).
\sq
%%%%%%%%%%%%%%%%%%%%%
\subsection{The $q$-$E_7^{(1)}$ equation from representation $2$}\label{sec:eq_E7_T2}
Let us define the following composition of the Weyl group elements $s_0, \ldots, s_7, \pi$ as
%Under the composed action 
\begin{equation}\label{eq:aW_E7_T2_translation} 
\begin{array}l
T_2=\pi s_{56543765402345123406543765430}, 
\end{array}
\end{equation}
where $s_{i_0i_1\ldots i_n}:=s_{i_0}s_{i_1}\cdots s_{i_n}$.
Then the action $T_2^{[2]}$ of $T_2$ in representation $2$ is expressed as %the root variables
%$\alpha_i$ in (\ref{eq:aW_E7_T1_root}) is transformed by
\begin{equation}\label{eq:aW_E7_T2_translation2}
\begin{array}l
T_2^{[2]}=\pi^{[2]} s_{56543765402345123406543765430}^{[2]}, 
\end{array}
\end{equation}
and we get 
\begin{prop}
The action $T_2^{[2]}$ (\ref{eq:aW_E7_T2_translation2}) 
%on the root 
gives the following 
%birational action of $x$, $y$:
equation
\begin{equation}\label{eq:aW_E7_T2_br}
\begin{array}l
\ds\Big(x-\frac{h_2}{qh_1}{T_2^{[2]}}^{-1}(y)+\frac{h_2}{qx}\Big)\Big(x-y+\frac{h_2}{x}\Big)=\dfrac{\prod_{i=3}^8(x-e_i)}{x^2\prod_{i=1,2}(x-\frac{h_1}{e_i})},
\\[5mm]
\dfrac{\{{T_2^{[2]}}(x)-\frac{{T_2^{[2]}}(h_1)}{z}\}(x-\frac{h_2}{z})}{\{{T_2^{[2]}}(x)-{T_2^{[2]}}(\frac{h_2}{h_1})\frac{z}{q}\}(x-z)}=\dfrac{z^2\prod_{i=3}^8(\frac{h_2}{z}-e_i)}{(\frac{h_2}{z})^2\prod_{i=3}^8(z-e_i)} \quad  for \ y=z+\dfrac{h_2}{z}.
\end{array}
\end{equation}
Here the equation (\ref{eq:aW_E7_T2_br}) is equivalent to the $q$-$E_7^{(1)}$ equation (\ref{eq:ev_E7_T2_ev}), and it has the singular point configuration (i.e. configuration type $2$ for surface type $q$-$A_1^{(1)}$ in Figure \ref{fig:E7-2})
\begin{equation}\label{eq:aW_E7_T2_8p}
(x,y)=\Big(e_i, e_i+\dfrac{h_2}{e_i}\Big)_{i=3,\ldots, 8},\left(\frac{h_1}{e_i},\infty\right)_{i=1,2}.
\end{equation}
\end{prop}

\prf
Firstly, from the actions $s_i^{[2]}$, $\pi^{[2]}$ (\ref{eq:aW_E7_T2_para}), (\ref{eq:aW_E7_T2_vari}), the action $T_2^{[2]}$ (\ref{eq:aW_E7_T2_translation2}) can be rewritten into the equation (\ref{eq:aW_E7_T2_br}). 

Next, under the transformation
%\footnote{see \cite{KNY17} for elliptic, multiplicative and additive cases of the symmetry type $E_8^{(1)}$ (i.e $e$-$E_8^{(1)}$, $q$-$E_8^{(1)}$, $d$-$E_8^{(1)}$).}
\begin{equation}\label{eq:aW_E7_T2_ve}
\begin{array}l
\ds v_i=\frac{e_i}{\lambda}, \quad u=\frac{z}{\lambda} \quad \kappa_i=\frac{h_i}{\lambda^2}, \quad f=\frac{x}{\lambda}, \quad g=\frac{y}{\lambda}, 
\end{array}
\end{equation}
where $T_2^{[2]}(h_1)=\frac{qe_1e_2h_2}{h_1}$, ${T_2^{[2]}}^{-1}(h_2)=\frac{qh_1^2}{h_2}$, $T_2^{[2]}(\lambda)=\frac{\lambda T_2^{[2]}(h_1)}{h_2}$, 
the equations (\ref{eq:aW_E7_T2_br}) is equivalent to the equation (\ref{eq:ev_E7_T2_ev}), and it has the configuration (\ref{eq:aW_E7_T2_8p}).
\sq
%%%%%%%%%%%%%%%%%%%%%%%%%%%%%%%%%%%%%%%%%%%%
\subsection{The $q$-$E_7^{(1)}$ equation from representation $3$}\label{sec:eq_E7_T3}
Let us define the following composition of the Weyl group elements $s_0, \ldots, s_7$ as
%Under the composed action 
\begin{equation}\label{eq:aW_E7_T3_translation} 
\begin{array}l
T_3=(s_{2 6 7 5 6 4 5 6 7 3 4 5 6 7 1 0 4 5 6 7 3 4 5 6 0 4 5 3 4})^2, 
\end{array}
\end{equation}
where $s_{i_0i_1\ldots i_n}:=s_{i_0}s_{i_1}\cdots s_{i_n}$.
Then the action $T_3^{[3]}$ of $T_3$ in representation $3$ is expressed as %the root variables
%$\alpha_i$ in (\ref{eq:aW_E7_T1_root}) is transformed by
\begin{equation}\label{eq:aW_E7_T3_translation2}
\begin{array}l
T_3^{[3]}=(s_{2 6 7 5 6 4 5 6 7 3 4 5 6 7 1 0 4 5 6 7 3 4 5 6 0 4 5 3 4}^{[3]})^2, 
\end{array}
\end{equation}
and we obtain 
\begin{prop}
The action $T_3^{[3]}$ (\ref{eq:aW_E7_T3_translation2}) 
%on the root 
gives the following 
%birational action of $x$, $y$:
equation
\begin{equation}\label{eq:aW_E7_T3_br}
\begin{array}l
\dfrac{\{{T_3^{[3]}}^{-1}(y)-(z+\frac{{T_3^{[3]}}^{-1}(h_2)}{z})\}\{y-(\frac{h_1}{z}+\frac{h_2z}{h_1})\}}{\{{T_3^{[3]}}^{-1}(y)-(\frac{h_1}{z}+\frac{{T_3^{[3]}}^{-1}(h_2)z}{h_1})\}\{y-(z+\frac{h_2}{z})\}}=\dfrac{U(\frac{h_1}{z})}{U(z)} \quad for\ x=z+\dfrac{h_1}{z},
\\[5mm]
\dfrac{\{{T_3^{[3]}}(x)-(z+\frac{{T_3^{[3]}}(h_1)}{z})\}\{x-(\frac{h_2}{z}+\frac{h_1z}{h_2})\}}{\{{T_3^{[3]}}(x)-(\frac{h_2}{z}+\frac{{T_3^{[3]}}(h_1)z}{h_2})\}\{x-(z+\frac{h_1}{z})\}}=\dfrac{U(\frac{h_2}{z})}{U(z)} \quad for \ y=z+\dfrac{h_2}{z}. 
\end{array}
\end{equation}
Here the equation (\ref{eq:aW_E7_T3_br}) is equivalent to the $q$-$E_7^{(1)}$ equation (\ref{eq:ev_E7_T3_ev}), and it has the singular point configuration (i.e. configuration type $3$ for surface type $q$-$A_1^{(1)}$ in Figure \ref{fig:E7-3})
\begin{equation}\label{eq:aW_E7_T3_8p}
(x,y)=\Big(e_i+\dfrac{h_1}{e_i}, e_i+\dfrac{h_2}{e_i}\Big)_{i=3,\ldots,8}, \Big(\frac{1}{\varepsilon}, \frac{1-\sfrac{e_1e_2}{h_1}}{(1-\sfrac{e_1e_2}{h_2})\varepsilon}\Big)_2.
\end{equation}
\end{prop}

\prf
Firstly, from the actions $s_i^{[3]}$ (\ref{eq:aW_E7_T3_para}) and (\ref{eq:aW_E7_T3_vari}), the action $T_3^{[3]}$ (\ref{eq:aW_E7_T3_translation2}) can be rewritten into the equation (\ref{eq:aW_E7_T3_br}). 

Next, under the transformation
%\footnote{see \cite{KNY17} for elliptic, multiplicative and additive cases of the symmetry type $E_8^{(1)}$ (i.e $e$-$E_8^{(1)}$, $q$-$E_8^{(1)}$, $d$-$E_8^{(1)}$).}
\begin{equation}\label{eq:aW_E7_T3_ve}
\begin{array}l
\ds v_i=\frac{e_i}{\lambda}, \quad u=\frac{z}{\lambda} \quad \kappa_i=\frac{h_i}{\lambda^2}, \quad f=\frac{x}{\lambda}, \quad g=\frac{y}{\lambda},
\end{array}
\end{equation}
where  $T_3^{[3]}(h_1)=\frac{qh_2^2}{h_1}$, ${T_3^{[3]}}^{-1}(h_2)=\frac{qh_1^2}{h_2}$, $T_3^{[3]}(\lambda)=\frac{\lambda T_3^{[3]}(h_1)}{h_2}$,   
the equation (\ref{eq:aW_E7_T3_br}) is equivalent to the  equation (\ref{eq:ev_E7_T3_ev}),  and it has the configuration (\ref{eq:aW_E7_T3_8p}). \sq

\begin{prop}\label{prop:aW_rel}
Two actions $T_1$ (\ref{eq:aW_E7_T1_translation}) and $T_3$ (\ref{eq:aW_E7_T3_translation})
%Two translations (\ref{eq:aW_E7_T1_translation}) and (\ref{eq:aW_E7_T3_translation})
 %on the root variables $\alpha_i$ (\ref{eq:aW_E7_T1_root}) and (\ref{eq:aW_E7_T3_root}) 
are transformed with each other under the following relation 
%\footnote{See Appendix \ref{sec:ev_rel} for the relation between (\ref{eq:ev_E7_T1_ev}) and (\ref{eq:ev_E7_T3_ev}) without using actions of affine Weyl group symmetry given in (\ref{eq:trans31}).} 
%written symbolically:
\begin{equation}\label{eq:trans31}
\begin{array}l
T_3=\psi^{-1} T_1\psi,
\end{array}
\end{equation}
where $\psi=s_{403423}$. 
\end{prop}
\prf 
Thanks to the fundamental relations (\ref{eq:aW_rel}), one can obtain the relation $T_3^{-1}\psi^{-1}T_1\psi=1$. \sq

%We note that
\begin{rem}
It seems that the action $T_2$ (\ref{eq:aW_E7_T2_translation}) can not be transformed into the actions $T_1$ (\ref{eq:aW_E7_T1_translation}) and $T_3$ (\ref{eq:aW_E7_T3_translation}) since the action $T_2$ contains the Dynkin diagram automorphism $\pi$. 
\end{rem}
\section{Conclusion}\label{sec:conc}
In this paper, we obtained the following main results.

\begin{itemize}
\item
We derived the three $q$-$E_7^{(1)}$ equaitons (\ref{eq:ev_E7_T1_ev}), (\ref{eq:ev_E7_T2_ev}),  (\ref{eq:ev_E7_T3_ev})
%\footnote{The equations (\ref{eq:ev_E7_T1_ev}) and   (\ref{eq:ev_E7_T2_ev}) have been known in \cite{GR99} and \cite{Nagao17-2, NY18-1, NY18-2} respectively.} 
by applying the degeneration through confluence of the $q$-$E_8^{(1)}$ equation in \cite{KNY17, Yamada14}. The $q$-$E_7^{(1)}$ equations (\ref{eq:ev_E7_T1_ev}), (\ref{eq:ev_E7_T2_ev}), (\ref{eq:ev_E7_T3_ev}) have different realizations (\ref{eq:ev_E7_T1_8p}), (\ref{eq:ev_E7_T2_8p}),  (\ref{eq:ev_E7_T3_8p}) (i.e. configuration types $1, 2, 3$ for surface type $q$-$A_1^{(1)}$ in Figures \ref{fig:E7-1}, \ref{fig:E7-2}, \ref{fig:E7-3}) in viewpoint of realizations of eight singular points configuration on the curve of bidgree $(2,2)$ in $\P^{1}\times\P^1$. %As far as we know, 
%We presented \red{the three kinds of realization in $\P^{1}\times\P^1$}. Both the realization (\ref{eq:ev_E7_T3_8p}) and the corresponding equation (\ref{eq:ev_E7_T3_ev}) are novel. %Furthermore we proved in Theorem \ref{prop:ev_rel_trans} that the evolution equations (\ref{eq:ev_E7_T1_ev}) and (\ref{eq:ev_E7_T3_ev}) is equivalent with each other under the relation (\ref{eq:ev_rel_E7_re1}) and (\ref{eq:ev_rel_E7_re2}).
\item
We gave the three representations $1, 2, 3$ 
%(\ref{eq:aW_E7_T1_para}), (\ref{eq:aW_E7_T1_vari}), (\ref{eq:aW_E7_T2_para}), (\ref{eq:aW_E7_T2_vari}), (\ref{eq:aW_E7_T3_para}), (\ref{eq:aW_E7_T3_vari})
 of affine Weyl  group actions with symmetry type $q$-$E_7^{(1)}$, and relations (\ref{eq:trans12}), (\ref{eq:trans23}) among the three representations $1, 2, 3$. The $q$-$E_7^{(1)}$ equations (\ref{eq:aW_E7_T1_br}), (\ref{eq:aW_E7_T2_br}), (\ref{eq:aW_E7_T3_br}) were derived from the actions (\ref{eq:aW_E7_T1_translation2}), (\ref{eq:aW_E7_T2_translation2}), (\ref{eq:aW_E7_T3_translation2}) of the three representations $1, 2, 3$ respectively. 
%affine Weyl group symmetry corresponding to the realizations (\ref{eq:aW_E7_T2_8p}) and (\ref{eq:aW_E7_T3_8p}). The relations %(\ref{eq:trans12}) and (\ref{eq:trans23}) 
%among the actions %(\ref{eq:aW_E7_T1_para}), (\ref{eq:aW_E7_T2_para}) and (\ref{eq:aW_E7_T3_para}) 
%were proved in Theorem \ref{thm:aW_rel}. Furthermore, it was showed in Corollary \ref{coro:aW_rel2} that the translations %(\ref{eq:aW_E7_T1_translation}) and (\ref{eq:aW_E7_T3_translation}) 
%are transformed with each other.
\end{itemize}
%\subsection{Summary}
%In this paper, extending the previous work \cite{NY16} to several variations, we obtained the following three main results.
%
%1. Choosing the proper variables according to several deformation directions, we obtained simply express the scalar Lax pairs, the evolution equations and particular solutions for the variations of the $q$-Garnier system in Theorems \ref{thm:Vn_NR_Gtrans} and \ref{thm:E7_ss_Gar_FG}.
%
%2. We also derived the reduced results from the above results in Theorems \ref{thm:Rn_NR_gtrans} and \ref{thm:E7_ss_E7_fg}).
%
%3. We clarified the relation among the $q$-Garnier system \cite{Sakai05-1, NY16}, Suzuki's system \cite{Suzuki17} and the $q$-KP hierarchy \cite{KNY02-2} in Theorem \ref{thm:SD_Dual} and Propositions \ref{prop:SD_Gar}, \ref{prop:SD_HOP}.

%\subsection{Problem}
%Several discrete Garnier systems have recently been studied in \cite{DST13, DT14, OR16-1, OR16-2}, except for \cite{Sakai05-1,Sakai05-2,NY16}. It seems interesting to construct particular solutions for these systems extending the results of \cite{Nagao16, NTY13, Yamada14}. 

\appendix
\section{Relation between $q$-$E_8^{(1)}$ equation (\ref{eq:ev_E8_ev}) and ORG from}\label{sec:ORG}

We give a correspondence between the $q$-$E_8^{(1)}$ equation given in \cite{KNY17, Yamada14} and ORG form \cite{ORG01}. The $q$-Painlev\'e equation with affine Weyl group symmetry of type $E_8^{(1)}$: 
\begin{equation}\label{eq:ORG_ev}
\begin{array}l
\dfrac{(\o{f}-g)(f-g)-(\frac{\kappa_1}{q}-\kappa_2)(\kappa_1-\kappa_2)\frac{1}{\kappa_2}}{(\frac{\o{f}q}{\kappa_1}-\frac{g}{\kappa_2})(\frac{f}{\kappa_1}-\frac{g}{\kappa_2})-(\frac{q}{\kappa_1}-\frac{1}{\kappa_2})(\frac{1}{\kappa_1}-\frac{1}{\kappa_2})\kappa_2}=\dfrac{\kappa_1^2}{q}\dfrac{A(\kappa_2,g)}{B(\kappa_2,g)}, \\[5mm]
\dfrac{(f-g)(f-\u{g})-(\kappa_1-\kappa_2)(\kappa_1-\frac{\kappa_2}{q})\frac{1}{\kappa_1}}{(\frac{f}{\kappa_1}-\frac{g}{\kappa_2})(\frac{f}{\kappa_1}-\frac{\u{g}q}{\kappa_2})-(\frac{1}{\kappa_1}-\frac{1}{\kappa_2})(\frac{1}{\kappa_1}-\frac{q}{\kappa_2})\kappa_2}=\dfrac{\kappa_2^2}{q}\dfrac{A(\kappa_1,f)}{B(\kappa_1,f)}, 
\end{array}
\end{equation}
were derived by Ohta-Rammani-Grammaticos in \cite{ORG01}.  
Here $A(h,x)$ and $B(h,x)$ are polynomials in $x$ of degree $4$ given by
\begin{equation}
\begin{array}l
A(h,x)=(\frac{m_0}{h^2}-\frac{m_2}{h}+m_4-hm_6+h^2m_8)+(\frac{m_1}{h^2}-m_5+2hm_7)x \\[5mm]
\phantom{A(h,x)}+(-\frac{m_0}{h^3}+m_6-3hm_8)x^2-m_7x^3+m_8x^4, 
\\[5mm]
B(h,x)=(\frac{m_0}{h^2}-\frac{m_2}{h^2}+m_4-hm_6+h^2m_8)+(\frac{2m_1}{h^2}-\frac{m_3}{h}+hm_7)x \\[5mm]
\phantom{B(h,x)}+(-\frac{3m_0}{h^3}+\frac{m_2}{h}-hm_8)x^2-\frac{m_1}{h^3}x^3+\frac{m_0}{h^4}x^4, 
\\[5mm]
U(z)=\frac{1}{z^4}\prod_{i=1}^8(z-u_i)=\frac{1}{z^4}\sum_{i=0}^8(-1)^im_iz^i.
\end{array}
\end{equation}
It is in \cite{Yamada14} shown that the equation (\ref{eq:ev_E8_ev}) is equivalent to the equation (\ref{eq:ORG_ev}) under the relation 
%the polynomials $A$ and $B$ are given by
\begin{equation}
A\Big(h,z+\frac{h}{z}\Big)=\frac{zU(z)-\frac{h}{z}U(\frac{h}{z})}{z-\frac{h}{z}}, \ B\Big(h,z+\frac{h}{z}\Big)=\frac{zU(\frac{h}{z})-\frac{h}{z}U(z)}{z-\frac{h}{z}}.   
\end{equation}

\section{Direct derivation of the relation between $q$-$E_7^{(1)}$ equations (\ref{eq:ev_E7_T1_ev}) and (\ref{eq:ev_E7_T3_ev})}\label{sec:ev_rel}
%\subsection{The relation between (\ref{eq:ev_E7_T1_ev}) and (\ref{eq:ev_E7_T2_ev})}
%\subsection{The relation between (\ref{eq:ev_E7_T1_ev}) and (\ref{eq:ev_E7_T3_ev})}
The relation between two kinds of $q$-$E_7^{(1)}$ equation (\ref{eq:ev_E7_T1_ev}) and (\ref{eq:ev_E7_T3_ev}) was obtained in Proposition \ref{prop:aW_rel}. Here we give its another direct derivation without using the affine Weyl group symmetry.
% for the time evolution (\ref{eq:ev_E8_te})
%, and show a relation between these equations (\ref{eq:ev_E7_T1_ev}) and (\ref{eq:ev_E7_T3_ev}).
%without using actions of affine Weyl group symmetry given in Section \ref{sec:aW}. 
%In order to distinguish dependent variables $(f,g)$ in (\ref{eq:ev_E7_T1_ev}) and (\ref{eq:ev_E7_T3_ev}), 
Replacing dependent variables $(f, g)$ and parameters $(\kappa_1, \kappa_2, v_1, v_2, v_3, v_4$, $v_5, v_6, v_7, v_8)$ by $(x,\frac{1}{y})$ and $(\kappa_1, \kappa_2, \dfrac{\kappa_1\kappa_2}{qv_6v_7v_8}, v_3, v_4, v_5, \dfrac{\kappa_1\kappa_2}{v_3v_4v_5}, v_6, v_7, v_8)$, then the $q$-$E_7^{(1)}$ equation (\ref{eq:ev_E7_T1_ev}) is rewritten as  
\begin{equation}\label{eq:ev_rel_E7_T1_ev1}
\ds\frac{x-\frac{\kappa_1}{\kappa_2}y}{x-y}\frac{x-\frac{\kappa_1q}{\kappa_2}\u{y}}{x-\u{y}}=\frac{x-\frac{qv_6v_7v_8}{\kappa_2}}{x-\frac{\kappa_1\kappa_2}{v_3v_4v_5}}\frac{\prod_{i=3}^5(x-\frac{\kappa_1}{v_i})}{\prod_{i=6}^8(x-v_i)}, 
\end{equation}
\begin{equation}\label{eq:ev_rel_E7_T1_ev2}
\ds\frac{y-\frac{\kappa_2}{\kappa_1}x}{y-x}\frac{y-\frac{\kappa_2q}{\kappa_1}\o{x}}{y-\o{x}}=\frac{y-\frac{qv_6v_7v_8}{\kappa_1}}{y-\frac{\kappa_1\kappa_2}{v_3v_4v_5}}\frac{\prod_{i=3}^5(y-\frac{\kappa_2}{v_i})}{\prod_{i=6}^8(y-v_i)}.
\end{equation}
Here the replacement satisfies the constraint (\ref{eq:E8_Le_cons}). Replacing the variable $z$ by $u$,  the $q$-$E_7^{(1)}$ equation (\ref{eq:ev_E7_T3_ev}) is  rewritten as 
\begin{equation}\label{eq:ev_rel_E7_T3_ev1}
\ds\frac{\frac{\kappa_1}{u}+\frac{\kappa_2}{\kappa_1}u-g}{u+\frac{\kappa_2}{u}-g}\frac{\frac{\kappa_1}{u}+\frac{\kappa_2}{q\kappa_1}u-\u{g}}{u+\frac{\kappa_2}{qu}-\u{g}}=\frac{u^6}{\kappa_1^3}\prod_{i=3}^8\frac{\frac{\kappa_1}{u}-v_i}{u-v_i} \quad \mbox{for} \ f=u+\frac{\kappa_1}{u}, 
\end{equation}
\begin{equation}\label{eq:ev_rel_E7_T3_ev2}
\ds\frac{\frac{\kappa_2}{u}+\frac{\kappa_1}{\kappa_2}u-f}{u+\frac{\kappa_1}{u}-f}\frac{\frac{\kappa_2}{u}+\frac{\kappa_1}{q\kappa_2}u-\o{f}}{u+\frac{\kappa_1}{qu}-\o{f}}=\frac{u^6}{\kappa_2^3}\prod_{i=3}^8\frac{\frac{\kappa_2}{u}-v_i}{u-v_i} \quad \mbox{for} \ g=u+\frac{\kappa_2}{u}.
\end{equation}

In order to relate (\ref{eq:ev_rel_E7_T1_ev1}), (\ref{eq:ev_rel_E7_T1_ev2}) and (\ref{eq:ev_rel_E7_T3_ev1}), (\ref{eq:ev_rel_E7_T3_ev2}), 
we put 
\begin{equation}\label{eq:ev_rel_E7_re1}
\ds x+\frac{\kappa_1}{x}-f=\frac{(x-y)\prod_{i=3}^5(x-\frac{\kappa_1}{v_i})}{x(x-\frac{\kappa_1\kappa_2}{v_3v_4v_5})(x-\frac{\kappa_1}{\kappa_2}y)},
\end{equation}
\begin{equation}\label{eq:ev_rel_E7_re2}
\ds y+\frac{\kappa_2}{y}-g=\frac{(y-x)\prod_{i=3}^5(y-\frac{\kappa_2}{v_i})}{y(y-\frac{\kappa_1\kappa_2}{v_3v_4v_5})(y-\frac{\kappa_2}{\kappa_1}x)}.
\end{equation}

Then we get the following two Lemmas,  
\begin{lem}\label{lem:ev_rel_birational}
The relations (\ref{eq:ev_rel_E7_re1}) and (\ref{eq:ev_rel_E7_re2}) give the birational transformation: $(f,g) \leftrightarrow (x,y)$. 
\end{lem}
\prf
Obviously, the relations (\ref{eq:ev_rel_E7_re1}) and (\ref{eq:ev_rel_E7_re2}) are the rational transform $(x,y) \rightarrow (f,g)$. Conversely, 
using $\prod_{i=3}^5(u-v_i)=u^3-\mu_1u^2+\mu_2u-\mu_3$, the relations (\ref{eq:ev_rel_E7_re1}) and (\ref{eq:ev_rel_E7_re2}) are written as
\begin{equation}\label{eq:ev_rel_E7_re3}
\ds x=\kappa_1\frac{f^2\kappa_2-fg\kappa_1+f(\mu_3-\kappa_2\mu_1)+g(\kappa_1\mu_1-\mu_3)+(\kappa_1-\kappa_2)(\kappa_1-\mu_2)}{f^2\mu_3-fg\mu_3+f\kappa_1(\kappa_2-\mu_2)+g\kappa_1(\mu_2-\kappa_1)+(\kappa_1-\kappa_2)(\kappa_1\mu_1-\mu_3)},
\end{equation}
\begin{equation}\label{eq:ev_rel_E7_re4}
\ds y=\kappa_2\frac{g^2\kappa_1-fg\kappa_2+g(\mu_3-\kappa_1\mu_1)+f(\kappa_2\mu_1-\mu_3)+(\kappa_2-\kappa_1)(\kappa_2-\mu_2)}{g^2\mu_3-fg\mu_3+g\kappa_2(\kappa_1-\mu_2)+f\kappa_2(\mu_2-\kappa_2)+(\kappa_2-\kappa_1)(\kappa_2\mu_1-\mu_3)},  
\end{equation}
and they are the rational transform $(f,g) \rightarrow (x,y)$. \sq

%Then we have
\begin{lem}\label{lem:ev_rel_trans}
The birational transformations (\ref{eq:ev_rel_E7_re1}), (\ref{eq:ev_rel_E7_re2}) (or equivalently (\ref{eq:ev_rel_E7_re3}), (\ref{eq:ev_rel_E7_re4})) can be written as 
\begin{equation}\label{eq:ev_rel_E7_re5}
\ds\frac{x-\frac{\kappa_1}{u}}{x-u}\frac{u+\frac{\kappa_2}{u}-g}{\frac{\kappa_1}{u}+\frac{\kappa_2}{\kappa_1}u-g}=\frac{\kappa_1^2}{u^4}\prod_{i=3}^5\frac{u-v_i}{\frac{\kappa_1}{u}-v_i}\quad \mbox{for}\quad f=u+\frac{\kappa_1}{u},
\end{equation}
\begin{equation}\label{eq:ev_rel_E7_re6}
\ds\frac{y-\frac{\kappa_2}{u}}{y-u}\frac{u+\frac{\kappa_1}{u}-f}{\frac{\kappa_2}{u}+\frac{\kappa_1}{\kappa_2}u-f}=\frac{\kappa_2^2}{u^4}\prod_{i=3}^5\frac{u-v_i}{\frac{\kappa_2}{u}-v_i}\quad \mbox{for}\quad g=u+\frac{\kappa_2}{u}.
\end{equation}
\end{lem}
\prf
Under a condition $f=u+\frac{\kappa_1}{u}$, eliminating $y$ from the relations (\ref{eq:ev_rel_E7_re1}) and (\ref{eq:ev_rel_E7_re2}) (i.e. (\ref{eq:ev_rel_E7_re1}) and (\ref{eq:ev_rel_E7_re4})), we have the relation (\ref{eq:ev_rel_E7_re5}). Under a condition $g=u+\frac{\kappa_2}{u}$, eliminating $x$ from the relations (\ref{eq:ev_rel_E7_re1}) and (\ref{eq:ev_rel_E7_re2}) (i.e. (\ref{eq:ev_rel_E7_re2}) and (\ref{eq:ev_rel_E7_re3})), we have the relation (\ref{eq:ev_rel_E7_re6}). \sq
\\

Thanks to Lemmas \ref{lem:ev_rel_birational} and \ref{lem:ev_rel_trans}, we have
\begin{prop}\label{prop:ev_rel_trans}
The equations (\ref{eq:ev_rel_E7_T1_ev1}), (\ref{eq:ev_rel_E7_T1_ev2}) and the equations (\ref{eq:ev_rel_E7_T3_ev1}), (\ref{eq:ev_rel_E7_T3_ev2}) are equivalent with each other under the relations\footnote{The relations (\ref{eq:ev_rel_E7_re1}) and (\ref{eq:ev_rel_E7_re2}) corresponds to the relation (\ref{eq:trans31}) in viewpoint of affine Weyl group symmetry.} (\ref{eq:ev_rel_E7_re1}) and (\ref{eq:ev_rel_E7_re2})
\end{prop}
%\noindent
%{\bf The proof of Proposition \ref{prop:ev_rel_trans}}
\prf
%We derive the equation (\ref{eq:ev_rel_E7_T3_ev2}) from  the equations (\ref{eq:ev_rel_E7_T1_ev2}), (\ref{eq:ev_rel_E7_T1_ev1}) and the relations (\ref{eq:ev_rel_E7_re1}), (\ref{eq:ev_rel_E7_re2}). 
Combining (\ref{eq:ev_rel_E7_T1_ev1}) and (\ref{eq:ev_rel_E7_re1}), we have 
\begin{equation}\label{eq:ev_rel_E7_re7}
\ds x+\frac{\kappa_1}{x}-f=\frac{(x-\frac{q\kappa_1}{\kappa_2}\u{y})\prod_{i=6}^8(x-v_i)}{x(x-\u{y})(x-\frac{qv_6v_7v_8}{\kappa_2})}.
\end{equation}
On the other hand, using (\ref{eq:ev_rel_E7_T1_ev2}) and (\ref{eq:ev_rel_E7_re2}), one has
\begin{equation}\label{eq:ev_rel_E7_re8}
\ds y+\frac{\kappa_2}{y}-g=\frac{(y-\frac{q\kappa_2}{\kappa_1}\o{x})\prod_{i=6}^8(y-v_i)}{y(y-\o{x})(y-\frac{qv_6v_7v_8}{\kappa_1})}.
\end{equation}
Substituting (\ref{eq:ev_rel_E7_re7}) into $\u{y}$ in $\u{(\ref{eq:ev_rel_E7_re8})}$, we get
\begin{equation}\label{eq:ev_rel_E7_re9}
\ds \frac{u+\frac{\kappa_2}{qu}-\u{g}}{\frac{\kappa_1}{u}+\frac{\kappa_2}{q\kappa_1}u-\u{g}}=\frac{\kappa_1}{u^2}\frac{x-\frac{\kappa_1}{u}}{x-u}\prod_{i=6}^8\frac{u-v_i}{\frac{\kappa_1}{u}-v_i}\quad \mbox{for} \ f=u+\frac{\kappa_1}{u}.
\end{equation}
Therefore the equation (\ref{eq:ev_rel_E7_T3_ev1}) is derived from the relations (\ref{eq:ev_rel_E7_re5}) and (\ref{eq:ev_rel_E7_re9}). 

Similarly, substituting (\ref{eq:ev_rel_E7_re8}) into $\o{x}$ in $\o{(\ref{eq:ev_rel_E7_re7})}$, we obtain 
\begin{equation}\label{eq:ev_rel_E7_re10}
\ds \frac{u+\frac{\kappa_1}{qu}-\o{f}}{\frac{\kappa_2}{u}+\frac{\kappa_1}{q\kappa_2}u-\o{f}}=\frac{\kappa_2}{u^2}\frac{y-\frac{\kappa_2}{u}}{y-u}\prod_{i=6}^8\frac{u-v_i}{\frac{\kappa_2}{u}-v_i}\quad \mbox{for} \ g=u+\frac{\kappa_2}{u}.
\end{equation}
Therefore the equation (\ref{eq:ev_rel_E7_T3_ev2}) is derived from the relations (\ref{eq:ev_rel_E7_re6}) and (\ref{eq:ev_rel_E7_re10}). 
\sq
%%%%%%%%%%%%%%%%%%%%%%%%%%%%%%
\section{Three translations on the root variables}\label{sec:trans}
For the Dynkin diagram
\begin{equation}\label{eq:aW_E7_dynkin}
\setlength{\unitlength}{0.5mm}
\begin{picture}(140,25)(0,10)
\put(0,15){\circle{4}}\put(2,15){\line(1,0){16}}
\put(20,15){\circle{4}}\put(22,15){\line(1,0){16}}
\put(40,15){\circle{4}}\put(42,15){\line(1,0){16}}
\put(60,15){\circle{4}}\put(62,15){\line(1,0){16}}
\put(80,15){\circle{4}}\put(82,15){\line(1,0){16}}
\put(100,15){\circle{4}}\put(102,15){\line(1,0){16}}
\put(120,15){\circle{4}}
\put(60,35){\circle{4}}\put(60,17){\line(0,1){16}}
\put(-2,6){\small$\alpha_{1}$}
\put(18,6){\small$\alpha_{2}$}
\put(36,6){\small$\alpha_{3}$}
\put(58,6){\small$\alpha_{4}$}
\put(78,6){\small$\alpha_{5}$}
\put(98,6){\small$\alpha_{6}$}
\put(118,6){\small$\alpha_{7}$}
\put(64,33){\small$\alpha_{0}$}
\end{picture}
\quad \delta_0 \Longleftrightarrow \delta_1, 
\end{equation}
we consider a pair of root bases $\{\alpha_i\}$, $\{\delta_i\}$ as symmetry/surface type $q$-$E_7^{(1)}$/$q$-$A_1^{(1)}$. Then we give three representations of root bases $\{\alpha_i^{[m]}\}$, $\{\delta_i^{[m]}\}$ $(m=1, 2, 3)$ as symmetry/surface type $E_7^{(1)}$/$A_1^{(1)}$ associated with the point configurations (\ref{eq:aW_E7_T1_8p}), (\ref{eq:aW_E7_T2_8p}), (\ref{eq:aW_E7_T3_8p}). We also show how the translations  $T_1^{[1]}$ (\ref{eq:aW_E7_T1_translation2}), $T_2^{[2]}$ (\ref{eq:aW_E7_T2_translation2}), $T_3^{[3]}$  (\ref{eq:aW_E7_T3_translation2}) act on the root variables $\alpha_i^{[1]}$, $\alpha_i^{[2]}$, $\alpha_i^{[3]}$ explicitly. 

%actions $T_1^{[1]}$ (\ref{eq:aW_E7_T1_translation2}), $T_2^{[2]}$ (\ref{eq:aW_E7_T2_translation2}), $T_3^{[3]}$  (\ref{eq:aW_E7_T3_translation2}) are regarded as translations on
\subsection{Case of representation $1$}
The root bases for representation $1$ are
\begin{equation}\label{eq:aW_E7_T1_root}
\begin{array}{ll}
%q\mbox{-}
E_7^{(1)}: &\alpha_0^{[1]}= H_1-H_2, \ \alpha_1^{[1]}= E_2-E_1,\ \alpha_2^{[1]}= E_3-E_2, \  
\alpha_3^{[1]}= E_4-E_3, \\
&\alpha_4^{[1]}=H_2-E_4-E_5, \ \alpha_5^{[1]}= E_5-E_6, \ \alpha_6^{[1]}= E_6-E_7,\ \alpha_7^{[1]}=  
E_7-E_8, \\
%q\mbox{-}
A_1^{(1)}: &\delta_0^{[1]}=H_1+H_2-E_1-E_2-E_3-E_4, \ \delta_1^{[1]}=  
H_1+H_2-E_5-E_6-E_7-E_8. 
\end{array}
\end{equation}
%\end{rem}

This realization for symmetry/surface type $E_7^{(1)}$/$A_1^{(1)}$ are associated with the actions $s_i^{[1]}$, $\pi^{[1]}$/configuration (\ref{eq:aW_E7_T1_8p}) (i.e. configuration type $1$ in Figure \ref{fig:E7-1}).

\begin{prop}\label{prop:aW_E7_T1_translation} 
%Let us define the following compositions of the Weyl group element as
%%Under the composed action 
%\begin{equation}\label{eq:aW_E7_T1_translation} 
%\begin{array}l
%%T_1=(s_0^{(1)} s_{4567345623451234}^{(1)})^2,
%T_1=(s_{\red{0}4567345623451234})^2. 
%\end{array}
%\end{equation}
%Then the action $T_1^{(1)}$ of $T_1$ in representation $(1)$ is expressed as %the root variables
%%$\alpha_i$ in (\ref{eq:aW_E7_T1_root}) is transformed by
%\begin{equation}\label{eq:aW_E7_T1_translation2}
%\begin{array}l
%T_1^{(1)}=(s_{04567345623451234}^{(1)})^2, 
%\end{array}
%\end{equation} and 
The action $T_1^{[1]}$ (\ref{eq:aW_E7_T1_translation2}) on the root variables $\alpha_i^{[1]}$ in (\ref{eq:aW_E7_T1_root}) are given by the translation
\begin{equation}\label{eq:aW_E7_T1_translation3}
\begin{array}l
T_1^{[1]}(\alpha_0^{[1]})=q^2\alpha_0^{[1]}, \ \ T_1^{[1]}(\alpha_4^{[1]})=\frac{\alpha_4^{[1]}}{q}, \ \ T_1^{[1]}(\alpha_i^{[1]})=\alpha_i ^{[1]} \ \ (i=1,2,3,5,6,7), 
\end{array}
\end{equation}
\end{prop}

\prf 
 Reading the relation (\ref{eq:aW_E7_T1_root}) multiplicatively and applying (\ref{eq:aW_E7_T1_para}), we obtain the desired result (\ref{eq:aW_E7_T1_translation3}). 
\sq
%%%%%%%%%%%%%%%%%%%%%%%%%%%%%%%%%%%%
\subsection{Case of representation $2$}
The root bases for representation $2$ are
\begin{equation}\label{eq:aW_E7_T2_root}
\begin{array}{ll}
%q\mbox{-}
E_7^{(1)}:&\alpha_0^{[2]}= H_2-E_3-E_4, \ \alpha_1^{[2]}= E_2-E_1, \ \alpha_2^{[2]}=H_1  
-E_2-E_3, \ \alpha_3= E_3-E_4, \\
&\alpha_4^{[2]}=E_4-E_5, \ \alpha_5^{[2]}= E_5-E_6, \ \alpha_6^{[2]}= E_6-E_7, \ \alpha_7^{[2]}=  
E_7-E_8, \\
%q\mbox{-}
A_1^{(1)}:&\delta_0^{[2]}=2 H_1+H_2-E_3-E_4-E_5-E_6-E_7-E_8, \ \delta_1^{[2]}=  
H_2-E_1-E_2.
\end{array}
\end{equation}
%\end{rem}

This realization for symmetry/surface type $E_7^{(1)}$/$A_1^{(1)}$ are associated with the actions $s_i^{[2]}$, $\pi^{[2]}$/configuration (\ref{eq:aW_E7_T2_8p}) (i.e. configuration type $2$ in Figure \ref{fig:E7-2}).

\begin{prop}\label{prop:aW_E7_T2_translation} 
The action $T_2^{[2]}$ (\ref{eq:aW_E7_T2_translation2}) on the root variables $\alpha_i^{[2]}$ in (\ref{eq:aW_E7_T2_root}) are given by the translation
\begin{equation}\label{eq:aW_E7_T2_translation3}
\begin{array}l
T_2^{[2]}(\alpha_0^{[2]})=\frac{\alpha_0^{[2]}}{q}, \ \  T_2^{[2]}(\alpha_2^{[2]})=q\alpha_2^{[2]}, \ \ T_2^{[2]}(\alpha_i^{[2]})=\alpha_i^{[2]} \ \ (i=1,3,4,5,6,7). 
\end{array}
\end{equation}
%where $s_{i_0i_1\ldots i_n}^{(1)}:=s_{i_0}^{(1)}s_{i_1}^{(1)}\cdots s_{i_n}^{(1)}$.
\end{prop}

\prf 
 Reading the relation (\ref{eq:aW_E7_T2_root}) multiplicatively and applying (\ref{eq:aW_E7_T2_para}), we obtain the desired result (\ref{eq:aW_E7_T2_translation3}). 
\sq
%%%%%%%%%%%%%%%%%%%%%%%%%%%%%%%%%%%%
\subsection{Case of representation $3$}
The root bases for representation $3$ are
\begin{equation}\label{eq:aW_E7_T3_root}
\begin{array}{ll}
%q\mbox{-}
E_7^{(1)}:&\alpha_0^{[3]}= E_3-E_4,\ \alpha_1^{[3]}= H_1-E_1-E_2,\ \alpha_2^{[3]}= H_2-H_1,\  
\alpha_3^{[3]}= H_1-E_3-E_4, \\
&\alpha_4^{[3]}=E_4-E_5,\ \alpha_5^{[3]}= E_5-E_6,\ \alpha_6^{[3]}= E_6-E_7,\ \alpha_7^{[3]}=  
E_7-E_8, \\
%q\mbox{-}
A_1^{(1)}:&\delta_0^{[3]}=2H_1+2H_2-2E_2-E_3-E_4-E_5-E_6-E_7-E_8,\ \delta_1^{[3]}=  
E_2-E_1.
\end{array}
\end{equation}
%\end{rem}

This realization for symmetry/surface type $E_7^{(1)}$/$A_1^{(1)}$ are associated with the actions $s_i^{[3]}$, $\pi^{[3]}$/configuration (\ref{eq:aW_E7_T3_8p}) (i.e. configuration type $3$ in Figure \ref{fig:E7-3}).
\begin{prop}\label{prop:aW_E7_T3_translation} 
The action $T_3^{[3]}$ (\ref{eq:aW_E7_T3_translation2}) on the root variables $\alpha_i^{[3]}$ in (\ref{eq:aW_E7_T3_root}) are given by the translation
\begin{equation}\label{eq:aW_E7_T3_translation3}
\begin{array}l
T_3^{[3]}(\alpha_i^{[3]})=\frac{\alpha_i^{[3]}}{q}, (\mbox{i=1,3}) \ \  T_3^{[3]}(\alpha_2^{[3]})=q^2\alpha_2^{[3]}, \ \  T_3^{[3]}(\alpha_i^{[3]})=\alpha_i^{[3]} \ (i=0,2,4,5,6,7).
\end{array}
\end{equation}
%where $s_{i_0i_1\ldots i_n}^{(1)}:=s_{i_0}^{(1)}s_{i_1}^{(1)}\cdots s_{i_n}^{(1)}$.
\end{prop}

\prf 
 Reading the relation (\ref{eq:aW_E7_T3_root}) multiplicatively and applying (\ref{eq:aW_E7_T3_para}), we obtain the desired result (\ref{eq:aW_E7_T3_translation3}). 
\sq
%%%%%%%%%%%%%%%%%%%%%%%%%%%%%%%%%%
%\red{
%\begin{rem}\label{rem:ev_rel_Lax}
%It was in \cite[Remark 4.1]{NY18-2} shown that the Lax equations for each evolution equations (\ref{eq:ev_E7_T1_ev}) and (\ref{eq:ev_E7_T2_ev}) are equivalent under a relation between independent variables. 
%\end{rem}
%}

%\begin{rem}\label{rem:ev_rel_trans}
%The remaining evolution equation (\ref{eq:ev_E7_T2_ev}) can not be transformed into the evolution equations (\ref{eq:ev_E7_T3_ev}) and (\ref{eq:ev_E7_T1_ev}). 
%\end{rem}

%In next section, we will explain Theorem \ref{prop:ev_rel_trans} and Remark \ref{rem:ev_rel_trans} by using bi-rational representation of affine Weyl group symmetry.

%\section{Representation of $q$-$E_8^{(1)}$ system (\ref{eq:ev_E8_ev}) from Weyl group}

%%%%%%%%%%%%%%%%%
\vskip10mm
%\vfill
%\break
\noindent
{\bf Acknowledgment.} 
The authors shall be thankful to Professor Tetsu Masuda, Masatoshi Noumi, Hidetaka Sakai, for valuable discussions. 
%The author is also grateful to the referees for stimulating comments. 
This work is partially supported by JSPS KAKENHI (22H01116), JSPS KAKENHI (19K14579) and JSPS KAKENHI (23K03173).
%and Expenses Revitalizing Education and Research of Akashi College (?????).

%The author shall be grateful to Professor Y. Yamada for variable discussions on this research and his infinite encouragement. 
%The author shall thank Professors Kenji Kajiwara, Tetsu Masuda, Masatoshi Noumi, Hidetaka Sakai, Takao Suzuki and Teruhisa Tsuda for stimulating comments and kindhearted supports.


\vskip5mm

\begin{thebibliography}{A}
%\bibitem{Adler94}
%Adler V. \'E, {\it Nonlinear chains and Painlev\'e equations}, Phys. D {\bf 73} (1994), no. 4, 335--351
%\bibitem{Doliwa14}
%Doliwa A., {\it Non-cmmutative rational Yang-Baxter maps}, Lett. Math. Phys. {\bf 104}, (2014) 299--309.
%\bibitem{DS84}
%Drinfel′d V. G., and Sokolov V. V., {\it Lie algebras and equations of Korteweg-de Vries type, Current problems in mathematics}, Vol. {\bf 24}, Itogi Nauki i Tekhniki, Akad. Nauk SSSR Vsesoyuz. Inst. Nauchn. i Tekhn. Inform., Moscow, (1984), pp. 81--180 (Russian).
%\bibitem{DST13}
%Dzhamay A., Sakai H., and Takenawa T., {\it Discrete Hamiltonian Structure of Schlesinger Transformations}, arXiv:1302.2972 [math-ph].
%\bibitem{DT14}
%Dzhamay A., and Takenawa T., {\it Geometric Analysis of Reductions from Schlesinger transformations to difference Painlev\'e equations}, arXiv:1408.3778 [math-ph].
%\bibitem{Fuchs05}
%Fuchs R., {\it Sur quelques \'equations diff\'erentielles lin\'eaires du second ordre}, C. R. Acad. Sci. (Paris) {\bf 141}, (1905) 555--558.
%\bibitem{Fuchs07}
%Fuchs R., \"{U}ber lineare homogene Differentialgleichungen zweiter Ordnung mit drei im Endlichen gelegene wesentlich singul\"{a}ren Stellen. Math. Ann. {\bf 63}, (1907) 301--321.
%\bibitem{Gambier10}
%Gambier B., {\it Sur les \'{e}quations diff\'{e}retielles du second ordre et du premier degr\'{e} dont l’int\'{e}grale g\'{e}n\'{e}rale est \'{a} points critiques fixes.} Acta. Math. {\bf33}, (1910) 1--55.
%\bibitem{Garnier12}
%Garnier R., {\it Sur des \'{e}quations diff\'{e}rentielles du troisi\'{e}me ordre dont l’int\'{e}grale g\'{e}n\'{e}rale est uniforme et sur une classe d’\'{e}quations nouvelles d’ordre sup\'{e}rieur dont l’int\'{e}grale g\'{e}n\'{e}rale a ses points critiques fixes.} Ann. Sci. Ecole Norm. Super. {\bf29}, (1912) 1--126.
%\bibitem{Garnier17}
%Garnier R., {\it Etudes de l’int\'egrale g\'en\'erale de l’\'equation {\rm VI} de M. Painlev\'e dans le voisinage de ses singularit\'e transcendentes}, Ann. Sci. Ecole Norm. Sup. (3) {\bf 34}, (1917) 239--353.
%\bibitem{GaR90}
%Gasper G. and Rahman M., {\it Basic Hypergeometric Series}, Encyclopedia of Mathematics
%and Its Applications, Volume {\bf 35} Cambridge University Press, Cambridge (1990)
%\bibitem{GaR04}
%Gasper G., and Rahman M., {\it Basic Hypergeometric Series. With a foreword by Richard Askey. Second edition. Encyclopedia of Mathematics and its Applications}, {\bf 96}. Cambridge University Press Cambridge (2004). 
%\bibitem{GNPRS94}
%Grammaticos B., Nijhoff F.W., Papageorgiou V.G., Ramani A., and Satsuma J., {\it Linearization and solution of the discrete Painlev\'{e} III equation}, Phys. Lett. A {\bf185}, (1994), 446.
%\bibitem{GNR99}
%Grammaticos B., Nijhoff F.W., Ramani A., Discrete Painlev\'{e} equations, in The Painlev\'{e} Property: One Century Later, Editor R. Conte, CRM Ser. Math. Phys., Springer, New York, (1999), 413--516.
\bibitem{GR99}
Grammaticos B., Ramani A.: On a novel $q$-discrete analogue of the Painlev\'e {\rm VI} equation. Phys. Lett. A. {\bf257} (1999), 288--292. 
%\bibitem{GRO03}
%Grammaticos, B., Ramani, A., Ohta, Y.: {\it A Unified Description of the Asymmetric $q$-$P_{V}$ and $d$-$P_{IV}$ Equations and their Schlesinger Transformations}. J. Nonlinear Math. Phys. {\bf10}(2), 215--228 (2003).
%\bibitem{Hasegawa13}
%Hasegawa K., {\it Quantizing the discrete Painlev\'e {\r mVI} equation: the Lax formalism}, Lett. Math. Phys. {\bf 103}, (2013) 865--879.
%\bibitem{HK07}
%Hamamoto T., and Kajiwara K., {\it Hypergeometric solutions to the q-Painlev\'e equation of type $A_4^{(1)}$}, J. Phys. A: Math. Theor. {\bf 40} (2007) 12509--12524.
%\bibitem{Ikawa13}
%Ikawa Y., {\it Hypergeometric Solutions for the $q$-Painlev\'e Equation of Type $E_6^{(1)}$ by the Pad\'e method}, Letters in Mathematical Physics, Volume {\bf 103}, Issue 7,  (2013) 743--763.
%\bibitem{IKSY91}
%Iwasaki K., Kimura H., Shimomura S., and Yoshida M., {\it From Gauss to Painlev\'e: A modern theory of special
%functions}, Vieweg, Braunschweig (1991).
%\bibitem{Jacobi}
%Jacobi,C.G.J {\it \"{U}ber die Darstellung einer Reihe gegebner Werthe durch eine gebrochne rationale Function}. J. Reine Angew. Math. {\bf 30}, (1846) 127--156.
%\bibitem{JS96}
%Jimbo M. and Sakai H., {\it A $q$-analog of the sixth Painlev\'e equation}, Lett. Math. Phys. {\bf 38} (1996) 145--154.
%\bibitem{JM81}
%Jimbo M., Miwa T., and Ueno K., {\it Monodromy preserving deformation of linear ordinary differential equations with rational coefficients. I.} Physica {\bf2D} (1981) 306--52.
%%坂井氏、モノドロミー保存変形で引用
%\bibitem{JM81-2}
%Jimbo M., and Miwa T., {\it Monodromy preserving deformation of linear ordinary differential equations with rational coefficients. II.}Physica {\bf2D} (1981) 407-448.
%\bibitem{JM81-3}
%Jimbo M., and Miwa T., {\it Monodromy preserving deformation of linear ordinary differential equations with rational coefficients. III.}Physica {\bf4D} (1981) 26--46.
%\bibitem{KK03}
%Kajiwara K., and Kimura K., {\it On a q-difference Painlev\'e III equation, I.Derivation, symmetry and Riccati type solutions.} J. Nonlinear Math. Phys.{\bf 10} (2003) 86--102.
%\bibitem{KMNOY03}
%Kajiwara K., Masuda T., Noumi M., Ohta Y., Yamada Y., ${}_{10}E_{9}$ solution to the elliptic Painlev\'{e} equation, J. Phys. A: Math. Gen. {\bf 36} (2003), L263-L272, nlin.SI/0303032.
%\bibitem{KMNOY04}
%Kajiwara K., Masuda T, Noumi M., Ohta Y. and Yamada Y.,
%{\it Hypergeometric solutions to the $q$-Painlev\'e equations}, 
%Int. Math. Res. Not. {\bf 2004} (2004) 2497--2521.
%\bibitem{KMNOY05}
%Kajiwara K., Masuda T., Noumi M., Ohta Y. and Yamada Y., {\it Construction of hypergeometric solutions to the $q$-Painlev\'e equations}, Int. Math. Res. Not. {\bf 2004} (2005), 1439--14533.
%\bibitem{KN15}
%Kajiwara K., and Nakazono N., {\it Hypergeometric solutions to the symmetric $q$-Painlev\'e equations}, Int. Math. Res. Not. {\bf 4} (2015), 1101--1140.
%\bibitem{KNT11}
%Kajiwara K., Nakazono N., and Tsuda T., {\it Projective reduction of the discrete Painlev\'e system
%of type $(A_2+A_1)^{(1)}$}, Int. Math. Res. Not. {\bf 2011} (2011), 930--966.
%\bibitem{KNY01}
%Kajiwara K., Noumi M., and Yamada Y., {\it A study on the fourth $q$-Painlev\'e equation}, J. Phys. A: Math. Gen. {\bf 34} (2001) 8563--8581.
%\bibitem{KNY02-1}
%Kajiwara K., Noumi M., and Yamada Y.,
%{\it Discrete dynamical systems with W($A_{m-1}^{(1)} \times A_{n-1}^{(1)}$) symmetry}, Lett. Math. Phys. {\bf 60} (2002), Issue 3, 211--219.
%\bibitem{KNY02-2}
%Kajiwara K., Noumi M., and Yamada Y., 
%{\it $q$-Painlev\'e Systems Arising from $q$-KP Hierarchy}, Lett. Math. Phys. {\bf 62} (2002), Issue 3, 259--268.
\bibitem{KNY17}
Kajiwara K., Noumi M.,Yamada Y.:
{\it Geometric aspects of Painlev\'e equations}. J. Phys. A: Math. Theor. {\bf 50} (2017), 073001--073164 (Topical Review). 
%arXiv 1509.08186 [nlin.SI].
%\bibitem{KK07}
%Kakei S., and Kikuchi T., {\it The sixth Painlev\'e equation as similarity reduction of
%$\widehat{{\mathfrak {gl}}}_{3}$ generalized Drinfel'd-Sokolov hierarchy}, Lett. Math. Phys. {\bf 79} (2007), no. 3, 221--234.
%\bibitem{Kawakami15}
%Kawakami H., {\it Matrix Painlev\'e systems}, J. Math. Phys. {\bf 56} (2015) 033503.
%\bibitem{KTGR00}
%Kruskal M. D., Tamizhmani K. M., B. Grammaticos and Ramani A.,  {\it Asymmetric discrete Painlev\'{e} equations}, Regul. Chaot. Dyn. {\bf 5}, (2000) 273--281.
%\bibitem{KO84}
%Kimura H., and Okamoto K., {\it On the polynomial Hamiltonian structure of the Garnier systems}. J. Math. Pures Appl. {\bf 63} (1984), 129--146.
%\bibitem{KOTY03}
%Kuniba A., Okado M., Takagi T., and Yamada Y.,  {\it Geometric crystal and tropical R for $D_{n}^{(1)}$}, Int. Math. Res. Not. {\bf 5}, (2003) 2565--2620.
%\bibitem{Kuroki11}
%Kuroki G.,  {\it Quantum groups and quantization of Weyl group symmetries of Painlev\'e systems}, in "Exploring new structures and natural constructions in mathematical physics" Adv. Stud. Pure Math., {\bf 61}, (2011) 289--325.
%\bibitem{Magnus95}
%Magnus A., Painlev\'{e}-type differential equations for the recurrence coefficients of semi- classical orthogonal polynomials, J. Comput. Appl. Math. {\bf 57} (1995) 215-237.
%\bibitem{Mano12}
%Mano T., {\it Determinant formula for solutions of the Garnier system and Pad\'e approximation}.
%J. Phys. A: Math. Theor. {\bf 45} (2012), 135206--135219.
%\bibitem{MT14}
%Mano T., and Tsuda T.,
%{\it  Two approximation problems by Hermite and the Schlesinger transformations} (Japanese),  
%RIMS Kokyuroku Bessatsu {\bf B47} (2014), 77--86.
%\bibitem{MT17}
%Mano T., and Tsuda T., {\it Hermite-Pad\'e approximation, isomonodromic deformation and hypergeometric integral}. Math. Z. {\bf 285} (2017), no. 1-2, 397--431.
%\bibitem{Masuda03}
%Masuda T., {\it On the rational solutions of $q$-Painlev\'e {\rm V} equation}, Nagoya Math. J. {\bf 169} (2003), 119--143.
%\bibitem{Masuda09}
%Masuda T., {\it Hypergeometric $\tau$-functions of the $q$-Painlev\'e system of type $E^{(1)}_7$}, SIGMA {\bf 5} (2009), 035 (30pp).
%\bibitem{Masuda11}
%Masuda T., {\it Hypergeometric $\tau$-functions of the $q$-Painlev\'e system of type $E^{(1)}_8$}, Ramanujan J. {\bf 24} (2011) 1--31.
%\bibitem{Mumford84}
%Mumford D., {\it Tata Lectures on Theta, II}, Birkh\"auser, (1984).
%\bibitem{MSY03}
%Murata M., Sakai H., and Yoneda J., {\it Riccati solutions of discrete Painlev\'e equations with Weyl group symmetry of type $E_8^{(1)}$}, J. Math. Phys. {\bf 44} (2003) 1396--1414.
%\bibitem{Murata04}
%Murata M., {\it New expressions for discrete Painlev\'e equations}, Funkcial. Ekvac. {\bf 47} (2004) 291--305. 
%\bibitem{Murata09}
%Murata M., {\it Lax forms of the $q$-Painlev\'e equations}, J. Phys. A: Math. Theor. {\bf 42} (2009) 115201.
\bibitem{Nagao15}
Nagao H.: {\it The Pad\'e interpolation method applied to $q$-Painlev\'e equations}. Lett. Math. Phys. {\bf 105} (2015), Issue 4, 503--521.
%\bibitem{Nagao16}
%Nagao H., {\it  Lax pairs for additive difference Painlev\'e equations}, arXiv:1604.02530 [nlin.SI].
%\bibitem{Nagao17-1}
%Nagao H., {\it The Pad\'e interpolation method applied to $q$-Painlev\'e equations II (differential grid version)}, Lett. Math. Phys. {\bf 107} Issue 1 (2017), 107--127.
\bibitem{Nagao17-2}
Nagao H.: {\it A variation of the $q$-Painlev\'e system with affine Weyl group symmetry of type $E_7^{(1)}$}. SIGMA. {\bf 13} (2017), 92--109.
%arXiv:1706.10087 [nlin.SI].
%\bibitem{Nagao17-3}
%Nagao H., {\it Hypergeometric special solutions for $d$-Painlev\'e equations}, arXiv:1706.10101 [nlin.SI].
\bibitem{NY18-1}
Nagao H., Yamada Y.: {\it Study of $q$-Garnier system by Pad\'e method}. Funkcialaj Ekvacioj. {\bf 61} (2018),109--133.
\bibitem{NY18-2}
Nagao H., Yamada Y.: {\it Variations of the $q$-Garnier system}. J.Phys. A: Math. Theor. {\bf 51} (2018), 135204--135222. %arXiv:1710.03998 [nlin.SI].
%\bibitem{NY21-1}
%Nagao, H., and Yamada, Y.: ``{\it Pad\'e method for Painlev\'e equations}". Math. Phys. {\bf 42} (2021) 90 pages, 
%\bibitem{Nakazono10}
%Nakazono N., {\it Hypergeometric $tau$ Functions of the $q$-Painlev\'e Systems of Type $(A_2+A_1)^{(1)}$} SIGMA {\bf 6} (2010) 084 (16pp).
%\bibitem{NSKGR96}
%Nijhoff F., Satsuma J., Kajiwara K., Grammaticos B., and Ramani, A., {\it A study of the alternate discrete Painlev\'e II equation}, Inv. Probl. {\bf 12} (1996) 697--716.
%\bibitem{NTY13}
%Noumi M., Tsujimoto S., and Yamada Y., {\it Pad\'{e} Interpolation for Elliptic Painlev\'e Equation}, Springer {\bf 40} (2013) 463--482.
%\bibitem{NY98}
%Noumi M., and Yamada Y., {\it Higher order Painlev\'e equations of type $A^{(1)}$}, Funkcial. Ekvac. {\bf 41} (1998), no. 3, 483--503.
\bibitem{ORG01}
Ohta Y., Ramani A., Grammaticos B.: {\it An affine Weyl group approach to the eight-parameter discrete Painlev\'e equation}. J. Phys. {\bf A34} (2001), 10523--10532.
%パンルヴェの論文はいつ？
%\bibitem{Okamoto86}
%Okamoto K., {\it Isomonodromic deformation and Painlev\'e equations, and the Garnier system}. J. Fac. Sci. Univ. Tokyo Sect. IA Math. {\bf 33} (1986), 575--618
%\bibitem{OR16-1}
%Ormerod, C.M., and Rains E.M., {\it Commutation Relations and Discrete Garnier Systems}, SIGMA {\bf 12} (2016), 110, 50 pages. 
%\bibitem{OR16-2}
%Ormerod, C.M., and Rains E.M., {\it An elliptic Garnier system}, arXiv:1607.07831 [nlin.SI]. 
%\bibitem{Painleve00}
%Painlev\'{e} P., {\it Sur les \'{e}quations diff\'{e}retielles du second ordre dont l’int\'{e}grale g\'{e}n\'{e}rale et uniforme. Oeuvret. III}, pp. 187--271
%Painlev\'e, P., {\it M\'emoire sur les \'equations diff\'erentielles dont l'int\'egrale g\'en\'erale est uniforme}, Bull. Soc. Math. Phys. France {\bf 28} (1900) 201--261.
%\bibitem{Painleve02}
%Painlev\'e, P., {\it Sur les \'equations diff\'erentielles du second ordre et d'ordre sup\'erieur dont l'int\'egrale g\'en\'erale est uniforme}, Acta Math. {\bf 25} (1902) 1--85. 
%\bibitem{QRT89}
%Quispel G. R. W., Roberts J. A. G., and Thompson C. J., {\it Integrable mappings and soliton equations II}, Physica D {\bf 34} (1989), 183--192.
%\bibitem{RG96}
%Ramani A., and Grammaticos B., {\it Discrete Painlev\'e equations:  coalescences, limits and degeneracies}, Physica A
%{\bf 228} (1996), 150--159.
%\bibitem{RG09}
%Ramani A., and Grammaticos B., {\it The number of discrete Painlev\'e equations is infinite}, Phys. Lett. A {\bf 373} (2009), 3028--3041 
%\bibitem{RGH91}
%Ramani A., Grammaticos B. and Hietarinta J., {\it Discrete Versions of the Painlev\'e Equations} Phys. Rev.Lett. {\bf 67} (1991), 1829--1832
%\bibitem{RGO01}
%Ramani A., Grammaticos B., and Ohta Y., {\it A Geometrical Description of the Discrete Painlev\'e VI and V Equations}, Commun. Math. Phys. {\bf 217} (2001), 315--329.
%\bibitem{RGTT01}
%Ramani A., Grammaticos B., Tamizhmani T., and Tamizhmani K.M., Special Function Solutions of the Discrete Painlev\'{e} Equations, Comput. Math. Appl. {\bf42} (2001), no. 3--5, 603--614.
%\bibitem{ROSG98}
%Ramani A., Ohta Y., Satsuma J., and Grammaticos B., {\it Self-duality and Schlesinger chains for the asymmetric d-PII and q-PIII equations}, Commum. Math. Phys. {\bf 192} (1998), 67--76. 
%\bibitem{qRT}
%G.~R.~W.~quispel, J.~A.~G.~Roberts, C.~J.~Thompson,
%{\it Integrable mappings and soliton equations},
%Phys. Lett. {\bf A126} (1988) 419--421.
%\bibitem{Rui}
%S.~N.~M.~Ruijsenaars, 
%{\it First degree analytic difference equations and %integrable quantum systems},
%J. Math. Phys. {\bf 38} (1997) 1069--1146. 
%\bibitem{Rains}
%E.~Rains, 
%{\it Recurrences for elliptic hypergeometric integrals},
%Rokko Lectures in Mathematics {\bf 18} (2005) 183--199.
%\bibitem{Sakai98}
%Sakai H., {\it Casorati determinant solutions for the $q$-difference sixth Painlev\'e equations}, Nonlinearity {\bf 11} (1998) 823--833.
\bibitem{Sakai01}
Sakai H.: {\it Rational surfaces with affine root systems and geometry of the Painlev\'e equations}. Commun. Math. Phys. {\bf 220} (2001),165--221.
%\bibitem{Sakai05-1} Sakai H., {\it A $q$-analog of the Garnier system}, Funkcialaj Ekvacioj, {\bf 48} (2005), 273--297.
%\bibitem{Sakai05-2} Sakai H., {\it Hypergeometric Solution of $q$-Schlesinger System of Rank Two}, Lett. Math. Phys. 
%{\bf 73} (2005), 237--247.
%\bibitem{Sakai06}
%Sakai H., {\it Lax form of the $q$-Painlev\'e equation associated with the $A_2^{(1)}$ surface}, J. Phys. A: Math. Gen. {\bf 39} (2006) 12203.
%\bibitem{Sakai10}
%Sakai H., {\it Isomonodromic deformation and 4-dimensional Painlev\'e type equations}, UTMS
%{\bf 2010-17} (Univ. of Tokyo 2010).
%\bibitem{Sasano06}
%Sasano Y., {\it Higher order Painlev\'e equations of type $D^{(1)}$}, RIMS Koukyuroku {\bf 1473} (2006),143--163.
%\bibitem{Sasano08}
%Sasano Y., {\it Coupled Painlev\'e {\rm VI} systems in dimension four with affine Weyl group symmetry of type $D^{(1)}$. {\rm II}}, RIMS Kokyuroku Bessatsu {\bf B5} (2008), 137--152.
%\bibitem{Skl}
%Sklyanin E. K.,
%{\it Separation of variables --- new trends}.
%Quantum field theory, integrable models and beyond (Kyoto, 1994).
%Progr. Theoret. Phys. Suppl. {\bf 118} (1995), 35--60. 
%\bibitem{SklTake}
%Sklyanin E. K. and Takebe T.,
%{\it Separation of variables in the elliptic Gaudin model}, 
%Comm. Math. Phys. {\bf 204} (1999), 17--38.
%\bibitem{Spi}
%V.~P.~Spiridonov,
%{\it Essays on the theory of elliptic hypergeometric functions}, 
%Uspekhi Matematicheskikh Nauk {\bf 63} (2008) 3--72.
%{\it Classical elliptic hypergeometric functions and their applications},
%Rokko Lectures in Mathematics {\bf 18} (2005) 253--287.
%\bibitem{SZ}
%V.~Spiridonov and A.~Zhedanov,
%{\it Spectral transformation chains and some new biorthogonal rational functions}, 
%Commun. Math. Phys. {\bf 210} (2000) 49--83.
%\bibitem{Suzuki13}
%Suzuki T., {\it A class of higher order Painlev\'e systems arising from integrable hierarchies of type $A$}, AMS Contemp. Math. {\bf 593} (2013) 125--141.
%\bibitem{Suzuki14}
%Suzuki T., {\it Six-dimensional Painlev\'e systems and their particular solutions in terms of rigid systems}, J. Math. Phys. {\bf 55} (2014) 102902. 
%\bibitem{Suzuki15}
%Suzuki T., {\it A $q$-analogue of the Drinfeld-Sokolov hierarchy of type $A$ and $q$-Painlev\'e system}, AMS Contemp. Math. {\bf 651} (2015), 25--38.
%\bibitem{Suzuki17}
%Suzuki T., {\it A reformulation of the generalized $q$-Painlev\'e {\rm VI} system with $W(A_{2n+1}^{(1)}) symmetry$}, J. Integrable Syst. {\bf 2} (2017), 1--18.
%\bibitem{Takenawa03}
%Takenawa T., {\it Weyl group symmetry of type $D_5^{(1)}$ in the $q$-Painlev\'e {\rm V} equation}, Funkcial. Ekvac. {\bf 46} (2003), 173--186.
%\bibitem{Tsuda04}
%Tsuda T., {\it Integrable mappings via rational elliptic surfaces}, J. Phys. $\mathrm{A}$:Math. Gen. {\bf 37} (2004), 2721--2730.
%\bibitem{Tsuda10}
%Tsuda T., {\it On an integrable system of q-difference equations satisfied by the universal characters: its Lax formalism and an application to $q$-Painlev\'e equations}, Comm. Math. Phys. {\bf 293} (2010), 347--359.
%\bibitem{Tsuda14}
%Tsuda T., {\it UC hierarchy and monodromy preserving deformation}, J. reine angew. Math. {\bf 690} (2014), 1--34.
%\bibitem{Tsuji}
%S.~Tsujimoto,
%{\it Determinant solutions of the nonautonomous %discrete Toda equation associated with the %deautonomized discrete KP hierarchy},
%J. Syst. Sci. Complex. {\bf 23} (2010) 153--176.
%\bibitem{Yamada01}
%Yamada Y., {\it A birational representation of Weyl group, combinatorial R-matrix and discrete Toda equation}, in "Physics and combinatorics, 2000", World Sci. Publ., (2001), 305--319.
%\bibitem{Yamada09-1}
%Yamada Y., {\it Pad\'e method to Painlev\'e equations}, Funkcial. Ekvac., {\bf 52} (2009), 83--92.
%\bibitem{Yamada09-2} 
%Yamada Y., {\it A Lax formalism for the elliptic difference Painlev\'e equation}, SIGMA, {\bf 5} (2009), 042 (15pp).
\bibitem{Yamada11}
Yamada Y.: {\it Lax formalism for $q$-Painlev\'e equations with affine Weyl group symmetry of type $E^{(1)}_n$}. IMRN. {\bf 17} (2011), 3823--3838.
%\bibitem{Yamada13}
%Yamada Y., Talk at the Workshop "{\it The $q$-Painlev\'{e} equations arising from the $q$-interpolation problems}", 09 July 2013,  Isaac Newton Institute for Mathematical Sciences.
\bibitem{Yamada14}
Yamada Y.: {\it A simple expression for discrete Painlev\'e equations}. RIMS Kokyuroku Bessatsu. {\bf B47} (2014), 087--095.
\end{thebibliography}
\end{document}